\documentclass[a4paper,11pt]{article}
\pdfoutput=1 
\usepackage{aas_macros,braket}
\usepackage{jcappub,natbib} 
\usepackage[T1]{fontenc} 

\title{Angular dependence of primordial trispectra and CMB spectral distortions}

\author[a]{Maresuke Shiraishi,}
\author[b,c,d]{Nicola Bartolo,}
\author[b,c,d]{and Michele Liguori}

\affiliation[a]{Kavli Institute for the Physics and Mathematics of the Universe (Kavli IPMU, WPI), UTIAS, The University of Tokyo, Chiba, 277-8583, Japan}
\affiliation[b]{Dipartimento di Fisica e Astronomia ``G. Galilei'', Universit\`a degli Studi di Padova, via Marzolo 8, I-35131, Padova, Italy}
\affiliation[c]{INFN, Sezione di Padova, via Marzolo 8, I-35131, Padova, Italy}
\affiliation[d]{INAF-Osservatorio Astronomico di Padova, Vicolo dell'OSservatorio 5, I-35122 Padova, Italy}

\emailAdd{maresuke.shiraishi@ipmu.jp}
\emailAdd{nicola.bartolo@pd.infn.it}
\emailAdd{michele.liguori@pd.infn.it}

\abstract{
Under the presence of anisotropic sources in the inflationary era, the trispectrum of the primordial curvature perturbation has a very specific angular dependence between each wavevector that is distinguishable from the one encountered when only scalar fields are present, characterized by an angular dependence described by Legendre polynomials. We examine the imprints left by curvature trispectra on the $TT\mu$ bispectrum, generated by the correlation between temperature anisotropies (T) and chemical potential spectral distortions ($\mu$) of the Cosmic Microwave Background (CMB). Due to the angular dependence of the primordial signal, the corresponding $TT\mu$ bispectrum strongly differs in shape from $TT\mu$ sourced by the usual $g_{\rm NL}$ or $\tau_{\rm NL}$ local trispectra, enabling us to obtain an unbiased estimation. From a Fisher matrix analysis, we find that, in a cosmic-variance-limited (CVL) survey of $TT\mu$, a minimum detectable value of the quadrupolar Legendre coefficient is $d_2 \sim 0.01$, which is 4 orders of magnitude better than the best value attainable from the $TTTT$ CMB trispectrum. In the case of an anisotropic inflationary model with a $f(\phi)F^2$ interaction (coupling the inflaton field $\phi$ with a vector kinetic term $F^2$), the size of the curvature trispectrum is related to that of quadrupolar power spectrum asymmetry, $g_*$. In this case, a CVL measurement of $TT\mu$ makes it possible to measure $g_*$ down to $10^{-3}$.
}

\begin{document}



\maketitle
\flushbottom

\section{Introduction}

Measurements of higher-order correlators of the primordial curvature fluctuation can play a crucial role in understanding the initial conditions of our Universe. In the usual single-field slow-roll inflationary scenario, the induced curvature perturbation is a nearly Gaussian field, and all the statistical information is then confined to the 2-point correlator or the power spectrum \cite{Acquaviva:2002ud,Maldacena:2002vr}. In contrast, higher-order correlators, such as the bispectrum and the trispectrum, are direct indicators for non-Gaussianity (NG), and their presence indicates the evidence for, e.g., some other source fields or some nonlinear interactions. Detailed analyses of their features, such as the shape and the scale dependence, or tests of the consistency relations between $n$-point and $(n-1)$-point correlators, thus provide essential information to select observationally viable Early Universe models (see e.g., \cite{Bartolo:2004if,Komatsu:2010hc,Chen:2010xka,Ade:2013ydc,Ade:2015ava} and references therein for review).

Primordial higher-order correlation functions have been deeply investigated using observational data of the Cosmic Microwave Background (CMB) anisotropies \cite{Komatsu:2008hk,Komatsu:2010fb,Bennett:2012zja}. Recent analyses using {\it Planck} data give constraints on primordial NGs with nearly cosmic-variance-limited (CVL) level accuracy \cite{Ade:2013ydc,Feng:2015pva,Ade:2015ava}, as long as CMB temperature anisotropies are concerned. Higher-order correlators related to Large Scale Structure (e.g., \cite{Giannantonio:2011ya, Maartens:2012rh, Byun:2014cea, Raccanelli:2015oma}) or 21-cm fluctuations (e.g., \cite{Cooray:1999kg,Cooray:2004kt,Cooray:2008eb, Pillepich:2006fj,Munoz:2015eqa, Shimabukuro:2015iqa}) are expected as future NG observables.

This paper focuses on another observable that has been shown to be particularly promising to constrain primordial NG, namely the correlation between CMB temperature (T) fluctuations and CMB $\mu$-type chemical potential spectral distortions, induced by heat release due to diffusion of acoustic waves, at redshifts from $2 \times 10^6 $ to $5 \times 10^4$. $\mu$-distortions display a quadratic dependence on the primordial curvature perturbation, while the temperature depends linearly on it. The curvature bispectrum and trispectrum can therefore source $T\mu$ and $\mu\mu$ correlations, respectively \cite{Pajer:2012vz}. Detectability analyses, based on futuristic $\mu$-distortion anisotropy surveys, have been carried on for several theoretically-motivated NG templates \cite{Pajer:2012vz,Ganc:2012ae,Biagetti:2013sr, Miyamoto:2013oua, Kunze:2013uja, Ganc:2014wia, Ota:2014iva, Emami:2015xqa, Shiraishi:2015lma,Dimastrogiovanni:2016aul,Ota:2016mqd}. Observational constraints on the usual local NG parameters, $f_{\rm NL}$ and $\tau_{\rm NL}$, based on {\it Planck} estimates of $T\mu$ and $\mu\mu$, are already available \cite{Khatri:2015tla}. In \cite{Bartolo:2015fqz}, we recently analyzed $TT\mu$ as an observable which depends on the curvature trispectrum. Our main finding was that, contrary to $\mu \mu$, $TT \mu$ is sensitive not only to $\tau_{\rm NL}$ but also to the other local trispectrum parameter, $g_{\rm NL}$, potentially improving with respect to the constraints that can be obtained with the trispectrum of CMB anisotropies.

An important difference between $\mu\mu$ and $TT\mu$ lies in the number of degrees of freedom: the angular power spectrum of $\mu\mu$ depends on only one $\ell$ mode, while $TT\mu$ varies in 3D harmonic space $(\ell_1, \ell_2, \ell_3)$. It is therefore expected that $TT\mu$ is more sensitive to some details of the NG shapes and has an advantage in discriminating between different primordial trispectrum shapes. In this paper, we examine $TT\mu$ generated from curvature trispectra with angular dependence \cite{Shiraishi:2013oqa}, characterized by
\begin{eqnarray} 
  \Braket{\prod_{n=1}^4 \zeta_{{\bf k}_n}}
  &=&  (2\pi)^3 \delta^{(3)}\left(\sum_{n=1}^4 {\bf k}_n \right)
  \sum_{L} d_L \left[ {\cal P}_L(\hat{k}_1 \cdot \hat{k}_3) + {\cal P}_L(\hat{k}_1 \cdot \hat{k}_{12}) + {\cal P}_L(\hat{k}_3 \cdot \hat{k}_{12}) \right]   \nonumber \\ 
&& \times
P(k_1) P(k_3) P(k_{12}) + (23~{\rm perm})  ~, \label{eq:zeta4_def}
\end{eqnarray}
where ${\bf k}_{12} \equiv {\bf k}_1 + {\bf k}_2$, $P(k)$ denotes the power spectrum of the curvature perturbation, and ${\cal P}_L(x)$ are the Legendre polynomials. This exactly expresses the angular dependence arising from the presence of anisotropic sources,%
\footnote{In this paper, ``anisotropic sources'' mean objects sourcing a nontrivial angle dependence between each wavevector in the angle-averaged observables or the isotropized curvature correlators like Eqs.~\eqref{eq:zeta4_def} and \eqref{eq:zeta3_def}.}
such as primordial vector fields present during inflation (see e.g.~\cite{Dimastrogiovanni:2010sm,Soda:2012zm,Maleknejad:2012fw,Bartolo:2012sd,Naruko:2014bxa,Bartolo:2014hwa,Bartolo:2015dga}). In addition to $d_0$, a nonzero $d_2$ appears in inflationary models where the inflaton field couples to a vector field via a $f(\phi)F^2$ interaction \cite{Shiraishi:2013vja,Abolhasani:2013zya,Shiraishi:2013oqa,Rodriguez:2013cj} (note that, for the $L = 0$ case, Eq.~\eqref{eq:zeta4_def} is independent of any angle and hence equivalent to a $\tau_{\rm NL}$-type trispectrum, with the replacement $d_0 = \tau_{\rm NL} / 6$). The same model also predicts nonzero $c_0$ and $c_2$ in the curvature bispectrum template \cite{Shiraishi:2013vja}:  
\begin{eqnarray}
  \Braket{\prod_{n=1}^3 \zeta_{{\bf k}_n} }
  = (2\pi)^3 \delta^{(3)}\left(\sum_{n=1}^3 {\bf k}_n \right) 
  \sum_{L} c_L {\cal P}_L(\hat{k}_1 \cdot \hat{k}_2) 
  P(k_1) P(k_2) 
  + (2~{\rm perm})
  ~, \label{eq:zeta3_def}
\end{eqnarray}
where the $L=0$ case is equivalent to the usual local NG template and hence $c_0 = (6/5) f_{\rm NL}$.%
\footnote{
  Other examples that give rise to bispectra and trispectra shapes of the type described in Eqs.~(\ref{eq:zeta4_def}) and~(\ref{eq:zeta3_def}) are the so-called solid inflation models~\cite{Endlich:2012pz, Bartolo:2013msa, Endlich:2013jia, Bartolo:2014xfa}, which are based on a specific internal symmetry obeyed by the inflaton fields, and which, e.g., produce in the bispectrum $c_2 \gg c_0$. Recently a model with a $f(\phi) (F^2 + F\tilde{F})$ coupling has been proposed as the first example of an inflationary model where $c_1$ is generated~\cite{Bartolo:2015dga}. Large-scale non-helical and helical magnetic fields in the radiation-dominated era do also generate $c_0$, $c_2$ and $c_1$~\cite{Shiraishi:2012rm, Shiraishi:2012sn, Shiraishi:2013vja}. See \cite{Ashoorioon:2016lrg} for other possibilities of generating anisotropic NGs.} 
Later we show that $T\mu$ due to $c_L$ and $\mu\mu$ due to $d_L$ vanish except for $ L=0$, while $TT\mu$ due to $d_L$ becomes nonzero for any even $L$. This is due to the difference of number of degrees of freedom mentioned above. 

The structure of this paper follows that of previous papers about $TTT$ from $c_L$ \cite{Shiraishi:2013vja} and $TTTT$ from $d_L$ \cite{Shiraishi:2013oqa}. We start by computing $TT\mu$ using the flat-sky approximation, and see how the 3D ${\bf k}$-space angular dependence in Eq.~\eqref{eq:zeta4_def} is projected to the 2D $\boldsymbol{\ell}$ space. After that, we recompute $TT\mu$ in full-sky and show, both via visual inspection and by actually computing correlation coefficients, that $TT\mu$ from $ d_2$ has a very different shape compared to those induced by $d_0$ (or equivalently $\tau_{\rm NL}$) and $g_{\rm NL}$. We then forecast error bars with a Fisher matrix analyses, showing that $d_2 \sim 0.01$, which is 4 orders of magnitude below the smallest detectable value from $TTTT$, is accessible by a CVL measurement of $TT\mu$. Finally, we focus on the $f(\phi) F^2$ model. In this case, due to the model-dependent consistency relations, $c_{0, 2}$ and $d_{0, 2}$ are expressed in terms of the parameter of the quadrupolar power spectrum asymmetry, $g_*$ \cite{Shiraishi:2013vja,Shiraishi:2013oqa,Bartolo:2012sd}. The sensitivities to $d_{0, 2}$ tell that $g_* \sim 10^{-3}$ is, in principle, accessible by $TT\mu$, and the 1D correlators $T\mu$ and $\mu\mu$, could further improve the sensitivity to $g_*$.

This paper is organized as follows. In the next section we compute $TT\mu$ from $d_L$ on the flat-sky and full-sky basis, and discuss residual angular dependence projected on $\boldsymbol{\ell}$--space. In Sec.~\ref{sec:fisher} we analyze the sensitivity to $d_L$ and some related parameters, and estimate the correlation coefficients between each shape, by employing the Fisher matrix. Section~\ref{sec:conclusions} contains our conclusions.
 
\section{Angular dependence in the $TT\mu$ bispectrum}\label{sec:TTmu}

In this section, we analyze signatures of the angle-dependent curvature trispectrum \eqref{eq:zeta4_def} in the $TT\mu$ bispectrum. Before employing the exact full-sky expression, we start to see how the angular dependence in ${\bf k}$ space is projected to $\boldsymbol{\ell}$ space, by employing the flat-sky formalism.

We consider effects of $\mu$-type spectral distortions induced by heating due to damping of acoustic waves, at redshifts varying from $z_i \sim 2 \times 10^6 $ to $z_f \sim 5 \times 10^4$. The injected heat depends on the photon energy density, therefore induced $\mu$-distortion anisotropies depend quadratically on the primordial curvature perturbations. This is summarized in the following formula, obtained via line-of-sight integration \cite{Sunyaev:1970er, 1975SvA....18..413I, 1982A&A...107...39D, 1991A&A...246...49B, Hu:1994bz, Chluba:2011hw, Khatri:2011aj, Chluba:2012gq, Khatri:2012tv, Khatri:2012tw,1968ApJ...151..459S, 1970ApJ...162..815P, 1983MNRAS.202.1169K,Weinberg:2008zzc,Pajer:2012vz, Ganc:2012ae, Khatri:2015tla}:
\begin{eqnarray}
\mu(\hat{n}) \simeq 
\left[ \prod_{n=1}^2 
\int \frac{d^3 {\bf k}_n}{(2\pi)^3} 
\zeta_{{\bf k}_n} 
\right] 
\int d^3 {\bf k}_3 
\delta^{(3)}\left(\sum_{n=1}^3 {\bf k}_n \right) 
f(k_1, k_2, k_3)
e^{-i {\bf k}_3 \cdot \hat{n} x_{\rm ls}} ~,  \label{eq:mu_real}  
\end{eqnarray}
where $\hat{n} = (\sin\theta \cos\phi, \sin\theta \sin\phi, \cos\theta)$ is the line-of-sight direction and $x_{\rm ls}$ is the conformal distance to the last scattering surface. The transfer function from $\zeta$ to $\mu$ is determined by the diffusion scales $k_D(z)$ at three specific redshifts, $k_i \equiv k_D(z_i) \sim 12000 \, {\rm Mpc}^{-1}$, $k_f \equiv k_D(z_f) \sim 46 \, {\rm Mpc}^{-1}$ and $k_{\rm rec} \equiv k_D(z \sim 1100) \sim 0.15 \, {\rm Mpc}^{-1}$, reading \cite{Khatri:2015tla}
\begin{equation}
   f(k_1, k_2, k_3) \simeq \frac{9}{4} \left[ e^{-(k_1^2 + k_2^2)/ k_i^2} - e^{-(k_1^2 + k_2^2)/ k_f^2} \right] e^{-k_3^2 / k_{\rm rec}^2}    ~.
   \end{equation}
The harmonic expansion, $a_{\ell m}^\mu = \int d^2 \hat{n} Y_{\ell m}^*(\hat{n}) \mu(\hat{n})$, results in the full-sky expression:
\begin{eqnarray}
a_{\ell m}^\mu =
4\pi (-i)^{\ell} \left[ \prod_{n=1}^2 
\int \frac{d^3 {\bf k}_n}{(2\pi)^3} \zeta_{{\bf k}_n}
\right]
\int d^3 {\bf k}_3 \delta^{(3)}\left(\sum_{n=1}^3 {\bf k}_n \right) 
Y_{\ell m}^*(\hat{k}_3)
 j_{\ell}(k_3 x_{\rm ls}) f(k_1, k_2, k_3) . \label{eq:alm_mu_full}
\end{eqnarray}
On the other hand, in the small $\theta$ (or equivalently large $\ell$) limit, the line-of-sight vector can be approximately projected on a flat space as $\hat{n} \simeq (\theta\cos\phi, \theta\sin\phi, 1) \equiv (\Theta_x, \Theta_y ,1)$, and the flat-sky expansion $a_{\boldsymbol{\ell}}^{\mu} = \int d^2 \boldsymbol{\Theta} e^{-i \boldsymbol{\ell} \cdot \boldsymbol{\Theta}} \mu(\Theta_x, \Theta_y ,1)$, then becomes reasonable \cite{Zaldarriaga:1996xe}. Substituting Eq.~\eqref{eq:mu_real} into this yields 
\begin{equation}
  a_{\boldsymbol{\ell}}^{\mu} = 
 \left[ \prod_{n=1}^2 
\int \frac{d^3 {\bf k}_n}{(2\pi)^3} 
\zeta_{{\bf k}_n} 
\right] 
\int d^3 {\bf k}_3 
 \delta^{(3)}\left(\sum_{n=1}^3 {\bf k}_n \right) 
 (2\pi)^2 \delta^{(2)}({\bf k}_3^\parallel x_{\rm ls} + \boldsymbol{\ell} )
 e^{  -i k_{3z}x_{\rm ls}} f(k_1, k_2, k_3) ~, \label{eq:alm_mu_flat}
\end{equation}
where ${\bf k} \equiv ({\bf k}^\parallel, k_z)$ (respectively the wave-vector components parallel and perpendicular to the plane  orthogonal to the line-of-sight).

In the same way, we can derive the full-sky and flat-sky expression of CMB temperature anisotropies \cite{Zaldarriaga:1996xe,Shiraishi:2010sm,Shiraishi:2010kd}, reading  
\begin{eqnarray}
  a_{\ell m}^{T} &=& 
4\pi i^{\ell} \int \frac{d^3 {\bf k}}{(2\pi)^{3}}
 \zeta_{\bf k} {\cal T}_{\ell}(k) Y_{\ell m}^*(\hat{k}) 
  ~, \label{eq:alm_T_full} \\
  a_{\boldsymbol{\ell}}^{T} &=&
  \int \frac{d^3 {\bf k}}{(2\pi)^3} \zeta_{\bf k} \int_0^{\tau_0} d\tau  S_T(k,\tau) 
  (2\pi)^2 \delta^{(2)}({\bf k}^\parallel D - \boldsymbol{\ell} ) e^{i k_{z} D}  ~, \label{eq:alm_T_flat}
  \end{eqnarray}
where $\tau_0$ is the present conformal time, $S_T(k,\tau)$ is the scalar-mode source function for temperature fluctuations, $D \equiv \tau_0 - \tau$ and ${\cal T}_\ell(k) \equiv \int_0^{\tau_0} d\tau S_T(k,\tau) j_\ell(kD)$.


\subsection{Flat-sky expression}

Using the flat-sky expressions \eqref{eq:alm_mu_flat} and \eqref{eq:alm_T_flat}, $TT\mu$ generated from the curvature trispectrum can be written as
\begin{eqnarray}
 \Braket{a_{\boldsymbol{\ell}_1}^{T} a_{\boldsymbol{\ell}_2}^{T} a_{\boldsymbol{\ell}_3}^{\mu}} &=& 
 \left[ \prod_{n=1}^2 \int \frac{d^3 {\bf k}_n}{(2\pi)^3}  \int_0^{\tau_0} d\tau_n  S_T(k_n,\tau_n)
  (2\pi)^2 \delta^{(2)}({\bf k}_n^\parallel D_n - \boldsymbol{\ell}_n ) e^{i k_{n z} D_n} \int \frac{d^3 {\bf K}_{n}}{(2\pi)^3}  \right]  \nonumber \\
 && \times \int d^3 {\bf K}_3 
 \delta^{(3)}\left(\sum_{n=1}^3 {\bf K}_n \right) 
 (2\pi)^2 \delta^{(2)}({\bf K}_3^\parallel x_{\rm ls} + \boldsymbol{\ell}_3 )
 e^{- i K_{3z} x_{\rm ls}}  \nonumber \\
&&
\times f(K_1, K_2, K_3) \Braket{\zeta_{{\bf k}_1} \zeta_{{\bf k}_2} \zeta_{{\bf K}_{1}}  \zeta_{{\bf K}_{2}} } ~. 
\end{eqnarray}
Due to the filtering by $f(K_1, K_2, K_3)$, all configurations except squeezed ones, ($K_1 \simeq K_2 \gg K_3$), are suppressed in the integrals above, enabling to approximate $\delta^{(3)}\left({\bf K}_1 + {\bf K}_2 + {\bf K}_3 \right) \simeq \delta^{(3)}\left({\bf K}_1 + {\bf K}_2 \right)$. For $L = {\rm even}$, the squeezed-limit ($K_1 \simeq K_2 \gg K_{12}$) expression of Eq.~\eqref{eq:zeta4_def} becomes
\begin{eqnarray}
\Braket{\zeta_{{\bf k}_1} \zeta_{{\bf k}_2} \zeta_{{\bf K}_{1}}  \zeta_{{\bf K}_{2}}}_{L = \rm even}
&\simeq& (2\pi)^3 \delta^{(3)}\left({\bf k}_1 + {\bf k}_2 + {\bf K}_1 + {\bf K}_2 \right) \nonumber \\
&& \times \sum_{L = \rm even} 4  d_L
\left[ {\cal P}_L(\hat{k}_1 \cdot \hat{K}_1) + {\cal P}_L(\hat{k}_1 \cdot \hat{K}_{12}) + {\cal P}_L(\hat{K}_1 \cdot \hat{K}_{12}) \right]
\nonumber \\ 
&& \times   P(k_1) P(K_1) P(K_{12})
  + ({\bf k}_1 \leftrightarrow {\bf k}_2)~;
\label{eq:zeta4_Leven_sq}
\end{eqnarray}
thus, we can write $TT\mu$ from $d_L$ as 
\begin{eqnarray}
  \Braket{a_{\boldsymbol{\ell}_1}^{T} a_{\boldsymbol{\ell}_2}^{T} a_{\boldsymbol{\ell}_3}^{\mu}}_{L = \rm even} &\simeq& 
  \left[ \prod_{n=1}^2 \int \frac{d^3 {\bf k}_n}{(2\pi)^3}  \int_0^{\tau_0} d\tau_n  S_T(k_n,\tau_n)
  (2\pi)^2 \delta^{(2)}({\bf k}_n^\parallel D_n - \boldsymbol{\ell}_n ) e^{i k_{n z} D_n}  \right] 
  \nonumber \\
&&     
\times \int \frac{d^3 {\bf K}_{1}}{(2\pi)^3} \int d^3 {\bf K}_3 
  (2\pi)^2 \delta^{(2)}({\bf K}_3^\parallel x_{\rm ls} + \boldsymbol{\ell}_3 )
e^{- i K_{3z} x_{\rm ls}} 
  \nonumber \\ 
&&
\times f(K_1, K_1, K_3) \delta^{(3)}\left({\bf k}_1 + {\bf k}_2 - {\bf K}_3 \right)
 \nonumber \\ 
&&\times \sum_{L = \rm even} 4 d_L \left[ \left({\cal P}_L(\hat{k}_1 \cdot \hat{K}_1) + {\cal P}_L(\hat{k}_1 \cdot \hat{K}_3) + {\cal P}_L(\hat{K}_1 \cdot \hat{K}_3) \right) \right. \nonumber \\ 
  \nonumber \\
  && \left. \times   P(k_1) P(K_1) P(K_3)
    + ({\bf k}_1 \leftrightarrow {\bf k}_2) \right] ~. \label{eq:aT2amu_flat}
\end{eqnarray}
On the other hand, for $L = {\rm odd}$, $\Braket{\zeta_{{\bf k}_1} \zeta_{{\bf k}_2} \zeta_{{\bf K}_{1}}  \zeta_{{\bf K}_{2}}}$ vanishes in the squeezed limit and hence $\Braket{a_{\boldsymbol{\ell}_1}^{T} a_{\boldsymbol{\ell}_2}^{T} a_{\boldsymbol{\ell}_3}^{\mu}}_{L = \rm odd}$ also vanishes. Cleaning up the wavevector integrals in Eq.~\eqref{eq:aT2amu_flat} leads to 
\begin{eqnarray}
 \Braket{a_{\boldsymbol{\ell}_1}^{T} a_{\boldsymbol{\ell}_2}^{T} a_{\boldsymbol{\ell}_3}^{\mu}}_{L = \rm even}
    &\simeq& \left[ \prod_{n=1}^2 \int_{-\infty}^{\infty} \frac{dk_{nz}}{2\pi}  \int_0^{\tau_0} d\tau_n  S_T(k_n,\tau_n) D_n^{-2} e^{i k_{nz} D_n} \right] \int_0^\infty \frac{K_1^2 dK_1}{2\pi^2} \nonumber \\ 
&& \times \int_{-\infty}^{\infty} dK_{3z} 
  x_{\rm ls}^{-2} e^{- i K_{3z} x_{\rm ls}} f(K_1, K_1, K_3) \nonumber \\ 
  && \times
  (2\pi)^2  
  \delta^{(2)}\left(\frac{\boldsymbol{\ell}_1}{D_1} + \frac{\boldsymbol{\ell}_2}{D_2} + \frac{\boldsymbol{\ell}_3}{x_{\rm ls}} \right)
\delta\left(k_{1z} + k_{2z} - K_{3z} \right)
  \nonumber \\
  && \times \sum_{L = \rm even} 4 d_L  \left[ {\cal P}_L(\hat{k}_1 \cdot \hat{K}_3 ) P(k_1)  + ({\bf k}_1 \leftrightarrow {\bf k}_2) \right] P(K_1) P(K_3) \nonumber \\ 
&&  \times \begin{cases}
   3 &: L = 0  \\
   1 &: L \geq 2
    \end{cases} ~,
\end{eqnarray}
where ${\bf k}_1 = (\frac{\boldsymbol{\ell}_1}{D_1} , k_{1z})$, ${\bf k}_2 = (\frac{\boldsymbol{\ell}_2}{ D_2} , k_{2z})$ and ${\bf K}_3 = (-\frac{\boldsymbol{\ell}_3}{ x_{\rm ls}} , K_{3z})$. Note that the difference of the factors for $L = 0$ and $L \geq 2$ is due to the fact that  $\int d^2 \hat{K}_1 {\cal P}_L(\hat{k}_1 \cdot \hat{K}_1)$, $\int d^2 \hat{K}_1 {\cal P}_L(\hat{k}_2 \cdot \hat{K}_1)$ and $\int d^2 \hat{K}_1 {\cal P}_L(\hat{K}_1 \cdot \hat{K}_3)$ in Eq.~\eqref{eq:aT2amu_flat} vanish except for $L = 0$.

We are now interested in the very high-$\ell$ behavior. The modes with $k_{1} D_1 \simeq \ell_1$, $k_2 D_2 \simeq \ell_2$ and $K_3 x_{\rm ls} \simeq \ell_3$ then contribute dominantly to the wavenumber integrals; thus, we may drop very small terms, such as $\frac{k_{1z} D_1 }{ \ell_1}$, $\frac{k_{2z} D_2}{\ell_2} $ and $\frac{K_{3z} x_{\rm ls}}{ \ell_3}$ in $\hat{k}_1 \cdot \hat{K}_3$ and $\hat{k}_2 \cdot \hat{K}_3$, resulting in 
\begin{eqnarray}
  \hat{k}_1 \cdot \hat{K}_3 &=&
  \frac{- \hat{\ell}_1 \cdot \hat{\ell}_3 + \frac{k_{1z}D_1}{\ell_1}\frac{K_{3z} x_{\rm ls}}{\ell_3}}
       {\sqrt{1 + (\frac{k_{1z}D_1}{\ell_1})^2}  \sqrt{1 + (\frac{K_{3z} x_{\rm ls}}{\ell_3})^2} }
       \simeq - \hat{\ell}_1 \cdot \hat{\ell}_3 ~, \label{eq:k1k3tol1l3} \\
  \hat{k}_2 \cdot \hat{K}_3 &=&
  \frac{- \hat{\ell}_2 \cdot \hat{\ell}_3 + \frac{k_{2z}D_2}{\ell_2}
    \frac{K_{3z} x_{\rm ls}}{\ell_3}}
       {\sqrt{1 + (\frac{k_{2z}D_2}{\ell_2})^2}  \sqrt{1 + (\frac{K_{3z} x_{\rm ls}}{\ell_3})^2} }
       \simeq - \hat{\ell}_2 \cdot \hat{\ell}_3  ~. \label{eq:k2k3tol2l3}
\end{eqnarray}
With these and a further large-$\ell$ approximation on the delta functions:
\begin{equation}
  \delta^{(2)}\left(\frac{\boldsymbol{\ell}_1}{D_1} + \frac{\boldsymbol{\ell}_2}{D_2} + \frac{\boldsymbol{\ell}_3}{x_{\rm ls}} \right)
   \delta(k_{1z} + k_{2z} - K_{3z}) 
  \simeq
\int_{-\infty}^\infty \frac{dr}{2\pi} e^{i(k_{1z} + k_{2z} - K_{3z}) r}
  r^2 \delta^{(2)}\left( \boldsymbol{\ell}_1 + \boldsymbol{\ell}_2 + \boldsymbol{\ell}_3\right)~,
  \end{equation}
we finally arrive at $\Braket{a_{\boldsymbol{\ell}_1}^{T} a_{\boldsymbol{\ell}_2}^{T} a_{\boldsymbol{\ell}_3}^{\mu}}_{L = \rm even} = (2\pi)^2 \delta^{(2)}\left(\boldsymbol{\ell}_1 + \boldsymbol{\ell}_2 + \boldsymbol{\ell}_3 \right) \sum_{L = \rm even} b_{\boldsymbol{\ell}_1 \boldsymbol{\ell}_2 \boldsymbol{\ell}_3}^{TT\mu, L = {\rm even}}$ where
 \begin{eqnarray}
 b_{\boldsymbol{\ell}_1 \boldsymbol{\ell}_2 \boldsymbol{\ell}_3}^{TT\mu, L = {\rm even}}
  &\simeq& 9 d_L A_S \ln\left( \frac{k_i}{k_f} \right) 
  \int_{-\infty}^\infty r^2 dr
 \left[ {\cal P}_L(\hat{\ell}_1 \cdot \hat{\ell}_3 ) {\cal B}_{\ell_1}^T(r) {\cal A}_{\ell_2}^T(r) + (\boldsymbol{\ell}_1 \leftrightarrow \boldsymbol{\ell}_2) \right]
      {\cal W}_{\ell_3}(r) \nonumber \\ 
&&
      \times \begin{cases}
   3 &: L = 0  \\
   1 &: L \geq 2
    \end{cases},
      \label{eq:blllTTmu_flat_dL}
 \end{eqnarray}
with  
\begin{eqnarray}
  {\cal A}_\ell^T(r) &\equiv&  \int_0^{\tau_0} d\tau 
  \int_{\ell / D}^{\infty} \frac{dk}{2\pi} \frac{1}{\sqrt{1 - \left(\frac{\ell}{kD}\right)^2  }} S_T(k,\tau)
  \frac{2}{D^2} \cos\left[ k(r + D) \sqrt{1 - \left(\frac{\ell}{kD}\right)^2  } \right] 
  ~, \\
  {\cal B}_\ell^T(r) &\equiv&  \int_0^{\tau_0} d\tau 
  \int_{\ell / D}^{\infty} \frac{dk}{2\pi} \frac{P(k)}{\sqrt{1 - \left(\frac{\ell}{kD}\right)^2  }} S_T(k,\tau)
  \frac{2}{D^2} \cos\left[ k(r + D) \sqrt{1 - \left(\frac{\ell}{kD}\right)^2  } \right] 
  ~, \\
 {\cal W}_{\ell}(r) &\equiv& \int_{\ell / x_{\rm ls}}^{\infty} \frac{dk}{2\pi} \frac{P(k)}{\sqrt{1 - (\frac{\ell}{k x_{\rm ls}})^2}} 
\frac{2}{x_{\rm ls}^2}\cos\left[ k (r + x_{\rm ls}) \sqrt{1 - \left(\frac{\ell}{k x_{\rm ls}}\right)^2} \right] e^{-k^2 / k_{\rm rec}^2}  ~.
  \end{eqnarray}
For this derivation, we have parametrized the curvature power spectrum as $P(k) = 2\pi^2 A_S k^{-3}$ and dealt with the $K_1$ integral as $\int_0^\infty K_1^2 d K_1 P(K_1)  \left[ e^{-2 K_1^2/ k_i^2} - e^{-2 K_1^2/ k_f^2} \right] \simeq 2\pi^2 A_S \ln(k_i/k_f)$.

\begin{figure}[t!]
  \begin{tabular}{cc}
\begin{minipage}{0.5\hsize}
  \begin{center}
    \includegraphics[width=1\textwidth]{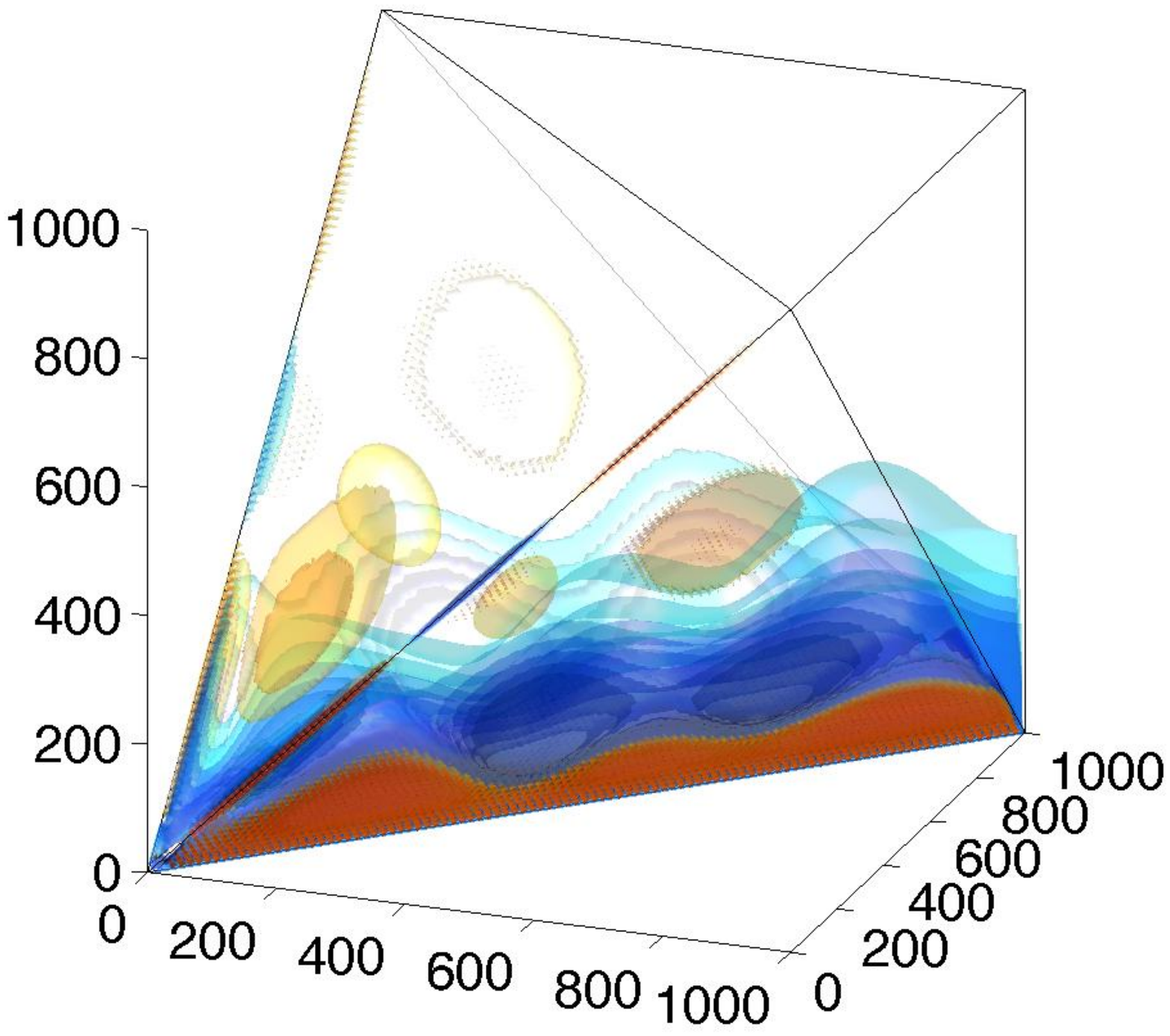}
    \hspace{1.6cm}$b_{\ell_1 \ell_2 \ell_3}^{TT\mu, d_2}$ (full-sky)
  \end{center}
\end{minipage}
\begin{minipage}{0.5\hsize}
  \begin{center}
    \includegraphics[width=1\textwidth]{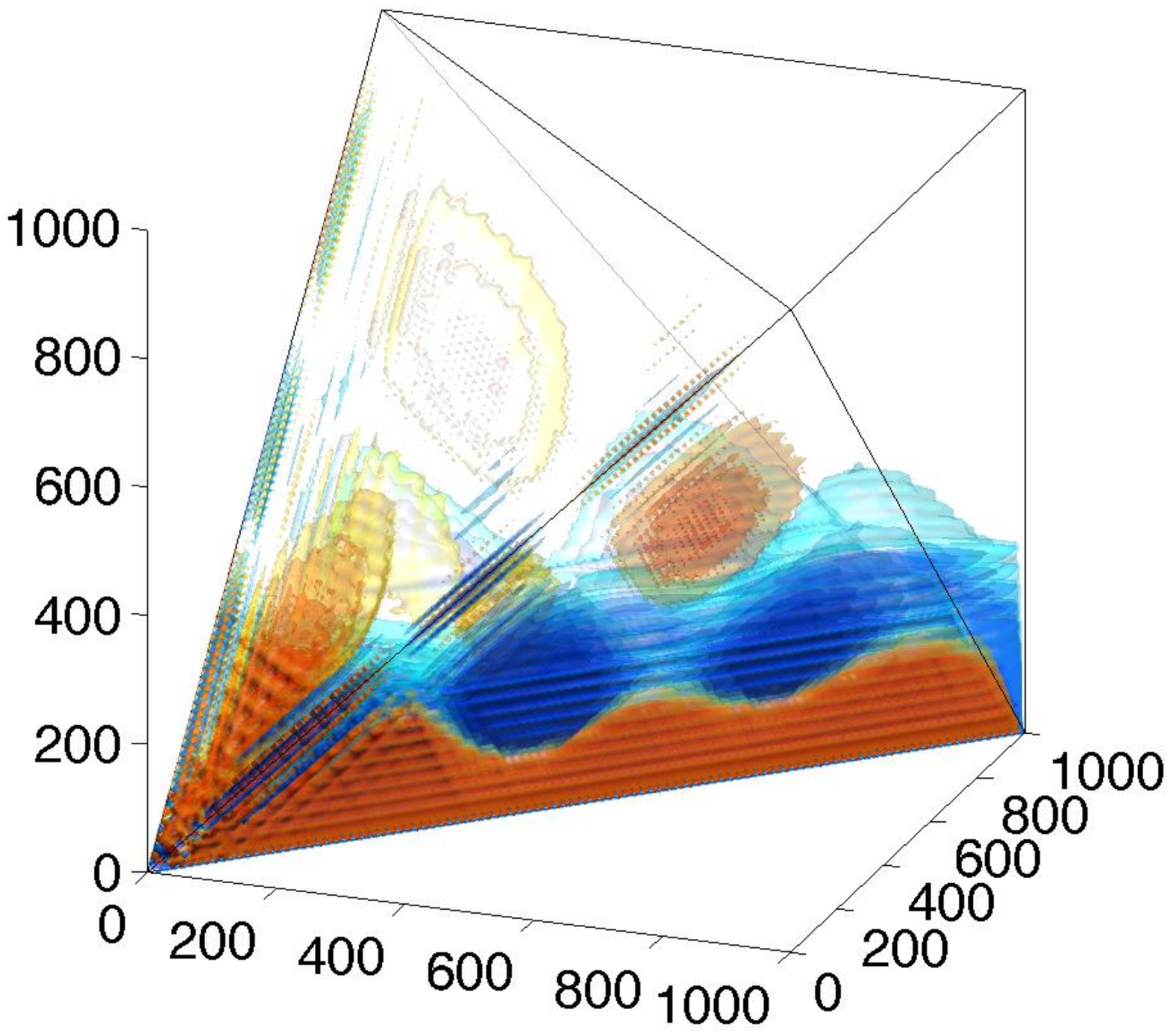}
    \hspace{1.6cm}$b_{\boldsymbol{\ell}_1 \boldsymbol{\ell}_2 \boldsymbol{\ell}_3}^{TT\mu, d_2}$ (flat-sky)
  \end{center}
\end{minipage}
  \end{tabular}
  \\
  \begin{tabular}{cc}
        \begin{minipage}{0.5\hsize}
  \begin{center}
    \includegraphics[width=1\textwidth]{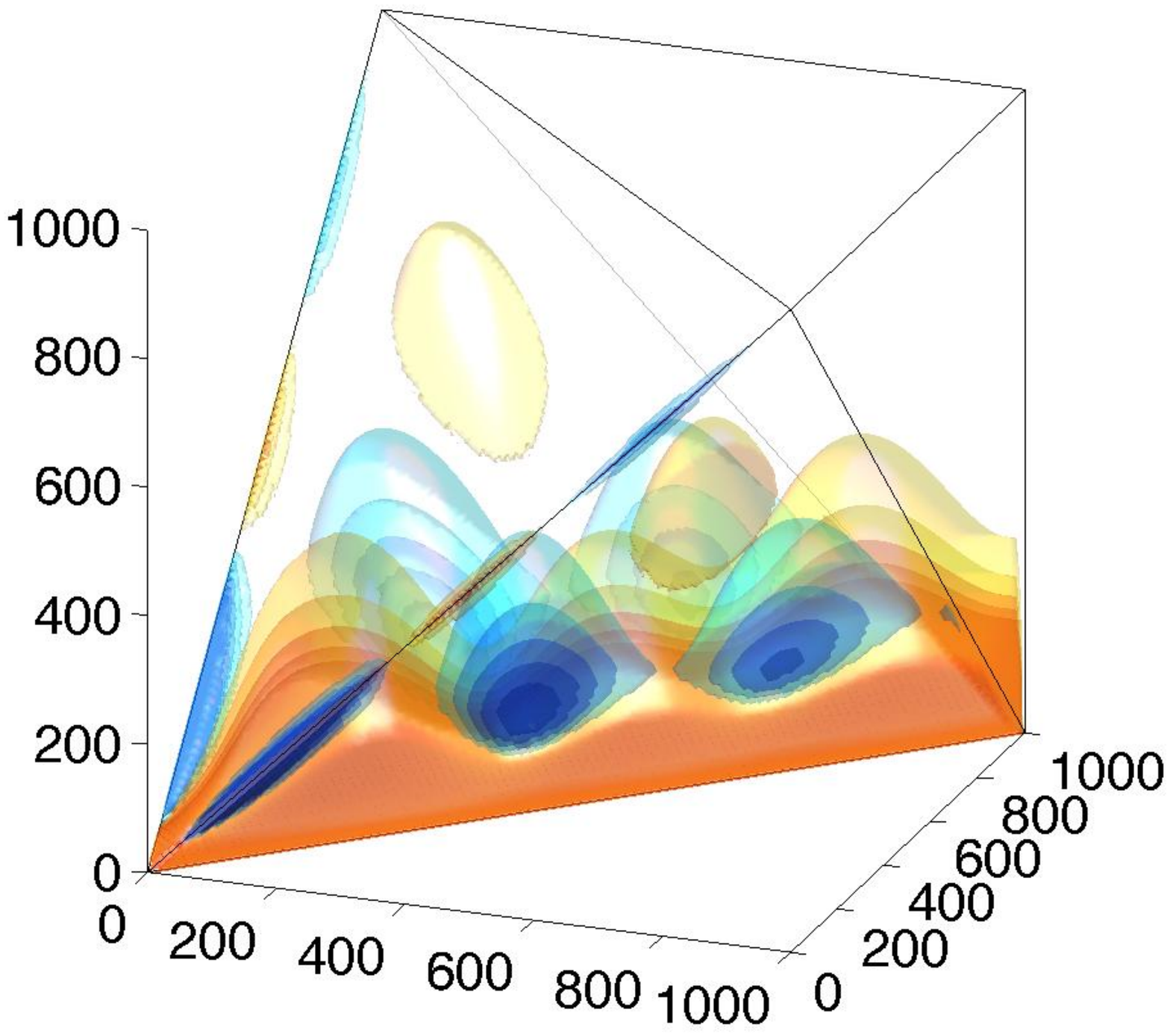}
    \hspace{1.6cm} $b_{\ell_1 \ell_2 \ell_3}^{TT\mu, d_0}$ or $b_{\ell_1 \ell_2 \ell_3}^{TT\mu, \tau_{\rm NL}}$
  \end{center}
\end{minipage}
    \begin{minipage}{0.5\hsize}
  \begin{center}
    \includegraphics[width = 1\textwidth]{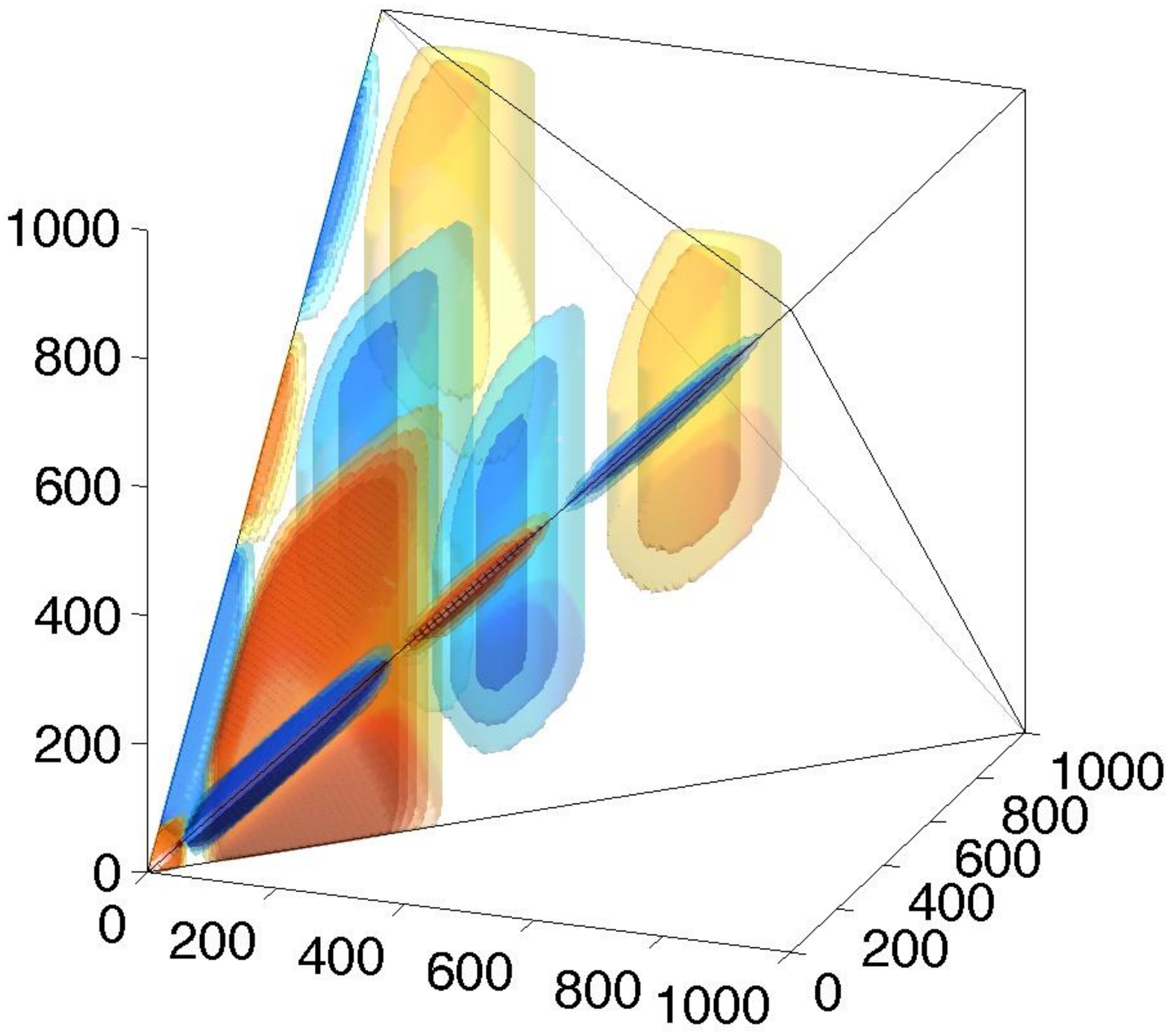}
    \hspace{1.6cm}$b_{\ell_1 \ell_2 \ell_3}^{TT\mu, g_{\rm NL}}$
  \end{center}
\end{minipage}
\end{tabular}
  \caption{3D representation of the $TT\mu$ bispectrum, $b_{\ell_1 \ell_2 \ell_3}^{TT\mu}$, normalized by the constant model template, $b_{\ell_1 \ell_2 \ell_3}^{TT\mu, \rm const} \propto \ell_1^{-5/4} \ell_2^{-5/4}$, derived in Appendix~\ref{appen:constmodel}. The vertical axis corresponds to $\ell_3$, associated with $\mu$, while the two bottom axes for $T$ correspond to $\ell_1$ and $\ell_2$. The left (right) top panel plots the full-sky (flat-sky) $b_{\ell_1 \ell_2 \ell_3}^{TT\mu, d_2}$, showing the strong resemblance between the full-sky and flat-sky expressions for $\ell_1, \ell_2, \ell_3 \gtrsim 100$. In the left (right) bottom panel we also describe the full-sky $b_{\ell_1 \ell_2 \ell_3}^{TT\mu, d_0}$ or equivalently $b_{\ell_1 \ell_2 \ell_3}^{TT\mu, \tau_{\rm NL}}$ ($b_{\ell_1 \ell_2 \ell_3}^{TT\mu, g_{\rm NL}}$). It is found from the left two panels that both $b_{\ell_1 \ell_2 \ell_3}^{TT\mu, d_0}$ and $b_{\ell_1 \ell_2 \ell_3}^{TT\mu, d_2}$ are mostly enhanced at $\ell_1 \sim \ell_2 \gg \ell_3$, however their color patterns differ from each other. The right bottom panel shows that $b_{\ell_1 \ell_2 \ell_3}^{TT\mu, g_{\rm NL}}$ is boosted at $\ell_1 \sim \ell_3 \gg \ell_2$ and $\ell_2 \sim \ell_3 \gg \ell_1$ and has quite different color distributions from $b_{\ell_1 \ell_2 \ell_3}^{TT\mu, d_0}$ or $b_{\ell_1 \ell_2 \ell_3}^{TT\mu, d_2}$.} \label{fig:blllTTmu_3D}
\end{figure}

It is obvious from the flat-sky expression \eqref{eq:blllTTmu_flat_dL} that the peculiar angular dependence of the curvature trispectrum \eqref{eq:zeta4_def} in ${\bf k}$ space is directly reflected on $\boldsymbol{\ell}$--space by the correspondence relations \eqref{eq:k1k3tol1l3} and \eqref{eq:k2k3tol2l3}. This results in a shape difference between $TT\mu$ bispectra for different $L$-modes. Equation~\eqref{eq:blllTTmu_flat_dL} allows us to derive the ratios of the $L=2$ bispectrum to the $L=0$ one for the squeezed-isosceles, flattened and equilateral triangles as
 \begin{eqnarray}
   \frac{b_{\boldsymbol{\ell}_1 \boldsymbol{\ell}_2 \boldsymbol{\ell}_3}^{TT\mu,L = 2}}{b_{\boldsymbol{\ell}_1 \boldsymbol{\ell}_2 \boldsymbol{\ell}_3}^{TT\mu , L = 0}}
   = \frac{d_2}{3 d_0} \times
   \begin{cases}
     -\frac{1}{2} &: \hat{\ell}_1 \cdot \hat{\ell}_3 = \hat{\ell}_2 \cdot \hat{\ell}_3  = 0 \ (\text{squeezed-isosceles})  \\
    1 &: |\hat{\ell}_1 \cdot \hat{\ell}_3| = |\hat{\ell}_2 \cdot \hat{\ell}_3|  = 1 \ (\text{flattened}) \\
    -\frac{1}{8} &: \hat{\ell}_1 \cdot \hat{\ell}_3 = \hat{\ell}_2 \cdot \hat{\ell}_3 = -\frac{1}{2} \ (\text{equilateral})
     \end{cases} ~, \label{eq:ratio_blllTTmu_L0_L2}
   \end{eqnarray}
 showing an example of size and sign changes of $ TT\mu$ due to the $L$-mode difference. The overall shape of $b_{\boldsymbol{\ell}_1 \boldsymbol{\ell}_2 \boldsymbol{\ell}_3}^{TT\mu,L = 2}$ is displayed in the right top panel of Fig.~\ref{fig:blllTTmu_3D}. The similarity of color pattern with the full-sky shape (left top panel of Fig.~\ref{fig:blllTTmu_3D}) confirms the accuracy of our flat-sky formula \eqref{eq:blllTTmu_flat_dL} for large $\ell$.
 
\subsection{Full-sky expression}

With Eqs.~\eqref{eq:alm_mu_full} and \eqref{eq:alm_T_full}, the full-sky $TT\mu$ is formulated as 
\begin{eqnarray}
\Braket{a_{\ell_1 m_1}^T a_{\ell_2 m_2}^T a_{\ell_3 m_3}^\mu}
  &=&  \left[\prod_{n=1}^2 4\pi i^{\ell_n} \int \frac{d^3 {\bf k}_n}{(2\pi)^{3}}
         {\cal T}_{\ell_n}(k_n)  Y_{\ell_n m_n}^*(\hat{k}_n) \int \frac{d^3 {\bf K}_n}{(2\pi)^3}  \right] \nonumber \\
         && \times 4\pi (-i)^{\ell_3} 
\int d^3 {\bf K}_3 \delta^{(3)}\left(\sum_{n=1}^3 {\bf K}_n \right) 
Y_{\ell_3 m_3}^*(\hat{K}_3)
j_{\ell_3}(K_3 x_{\rm ls}) \nonumber \\
&& \times f(K_1, K_2, K_3) 
\Braket{\zeta_{{\bf k}_1} \zeta_{{\bf k}_2} \zeta_{{\bf K}_1}  \zeta_{{\bf K}_2} } ~.
\end{eqnarray}
Also in this equation, the wavevector integrals are determined by the squeezed-limit signal ($K_1 \simeq K_2 \gg K_{12}$) of $\Braket{\zeta_{{\bf k}_1} \zeta_{{\bf k}_2} \zeta_{{\bf K}_1}  \zeta_{{\bf K}_2}}$. In the angle-dependent trispectrum case, such signal is highly suppressed for $L = {\rm odd}$, so a nonvanishing $TT\mu$ is realized only for $L = {\rm even}$. Plugging Eq.~\eqref{eq:zeta4_Leven_sq} into the above equation and evaluating the $\hat{K}_1$ and ${\bf K}_2$ integral with the squeezed-limit approximation: $\delta^{(3)}\left({\bf K}_1 + {\bf K}_2 + {\bf K}_3 \right) \simeq \delta^{(3)}\left({\bf K}_1 + {\bf K}_2 \right)$, yields
\begin{eqnarray}
 \Braket{a_{\ell_1 m_1}^T a_{\ell_2 m_2}^T a_{\ell_3 m_3}^\mu}_{L= \rm even}
  &\simeq& \left[\prod_{n=1}^2 4\pi i^{\ell_n} \int \frac{d^3 {\bf k}_n}{(2\pi)^{3}} {\cal T}_{\ell_n}(k_n)  Y_{\ell_n m_n}^*(\hat{k}_n) \right] \int_0^\infty \frac{K_1^2 dK_1}{2\pi^2}  \nonumber \\ 
         && \times 4\pi (-i)^{\ell_3}
\int d^3 {\bf K}_3 
Y_{\ell_3 m_3}^*(\hat{K}_3)
j_{\ell_3}(K_3 x_{\rm ls}) \nonumber \\
&& \times f(K_1, K_1, K_3) \delta^{(3)}\left({\bf k}_1 + {\bf k}_2 -{\bf K}_3 \right) \nonumber \\
&& \times \sum_{L = \rm even} 4 d_L \left[ {\cal P}_L(\hat{k}_1 \cdot \hat{K}_3) P(k_1) + ({\bf k}_1 \leftrightarrow {\bf k}_2) \right] P(K_1) P(K_3) \nonumber \\ 
&& \times 
\begin{cases}
   3 &: L = 0  \\
   1 &: L \geq 2
    \end{cases} ~.
  \end{eqnarray}
The angle dependence of the wavevectors in this equation can be decomposed using spherical harmonics, together with the identities:
\begin{eqnarray}
  {\cal P}_L(\hat{k}_1 \cdot \hat{k}_2) &=& \frac{4\pi}{2L+1} \sum_{M} Y_{LM}^*(\hat{k}_1) Y_{LM}(\hat{k}_2)   ~, \\
  \delta^{(3)}\left( \sum_{n=1}^3 {\bf k}_n \right) 
&=& 8 \int_0^\infty r^2 dr 
\left[ \prod_{n=1}^3 \sum_{L_n M_n} 
 j_{L_n}(k_n r) 
Y_{L_n M_n}^*(\hat{k}_n) \right] 
\nonumber \\
&&\times 
(-1)^{\frac{L_1 + L_2 + L_3}{2}}
h_{L_1 L_2 L_3}
\left(
  \begin{array}{ccc}
  L_1 & L_2 & L_3 \\
  M_1 & M_2 & M_3 
  \end{array}
  \right) ~,
\end{eqnarray}
with $h_{l_1 l_2 l_3} \equiv \sqrt{\frac{(2 l_1 + 1)(2 l_2 + 1)(2 l_3 + 1)}{4 \pi}} \left(
  \begin{array}{ccc}
  l_1 & l_2 & l_3 \\
   0 & 0 & 0 
  \end{array}
  \right)$. After expressing the angular integrals over products of spherical harmonics in terms of Wigner symbols, summing over angular momenta similarly as done in \cite{Shiraishi:2010kd,Shiraishi:2013vja,Bartolo:2015fqz}, and dealing with the $K_1$ integral in the same manner as for the previous flat-sky computation, we finally get the angle-averaged form
  \begin{equation}
  \Braket{a_{\ell_1 m_1}^T a_{\ell_2 m_2}^T a_{\ell_3 m_3}^\mu}_{L = \rm even}
  = h_{\ell_1 \ell_2 \ell_3}
\left(
  \begin{array}{ccc}
  \ell_1 & \ell_2 & \ell_3 \\
  m_1 & m_2 & m_3 
  \end{array}
  \right) \sum_{L = \rm even} b_{\ell_1 \ell_2 \ell_3}^{TT\mu, L = \rm even}
  \end{equation}
 with
 \begin{eqnarray}
  b_{\ell_1 \ell_2 \ell_3}^{TT\mu, L = \rm even}
  &\simeq& 9 d_L A_S \ln\left(\frac{k_i}{k_f}\right)
\frac{4\pi}{2L+1}  \sum_{L_1 L_3}
(-1)^{\frac{L_1 + L_3 + \ell_1 + \ell_3}{2}}
\frac{h_{L_1 \ell_2 L_3}}{h_{\ell_1 \ell_2 \ell_3}}
h_{\ell_1 L_1 L}h_{\ell_3 L_3 L}
  \left\{
  \begin{array}{ccc}
  \ell_1 & \ell_2 & \ell_3 \\
  L_3 & L & L_1 
  \end{array}
  \right\} 
 \nonumber \\ 
 && \times \int_0^\infty r^2 dr \beta_{\ell_1 L_1}^T(r) \alpha_{\ell_2}^T(r) \omega_{\ell_3 L_3}(r) \times
 \begin{cases}
   3 &: L = 0  \\
   1 &: L \geq 2
    \end{cases}
 \, + \, (\ell_1 \leftrightarrow \ell_2) ~, \label{eq:blllTTmu_all_dL}
\end{eqnarray}
and 
\begin{eqnarray}
 \alpha_{\ell}^T(r)
  &\equiv& \frac{2}{\pi} \int_0^\infty k^2 d k {\cal T}_{\ell}(k) j_{\ell}(k r) ~, \\
  \beta_{\ell L}^T(r)
  &\equiv& \frac{2}{\pi} \int_0^\infty k^2 d k P(k) {\cal T}_{\ell}(k) j_{L}(k r) ~, \\
   \omega_{\ell L}(r) &\equiv& \frac{2}{\pi} \int_0^\infty  k^2 dk P(k)  j_{\ell}(k x_{\rm ls}) j_{L}(k r) e^{-k^2 / k_{\rm rec}^2} ~.
\end{eqnarray}
In the small-$\ell$ limit, the Sachs-Wolfe (SW) approximation: ${\cal T}_\ell(k) \approx  -\frac{1}{5}j_\ell(k x_{\rm ls}) $, is reasonable and hence
\begin{eqnarray}
  \alpha_\ell^T(r) &\approx& -\frac{\delta(r - x_{\rm ls})}{5 x_{\rm ls}^2} ~, \\
 \beta_{\ell L}^T(x_{\rm ls}) &\approx&
  - \frac{\pi^2 }{10} A_S \frac{ \Gamma(\frac{\ell+L}{2})}
{\Gamma(\frac{\ell-L+3}{2}) \Gamma(\frac{L-\ell+3}{2})
\Gamma(\frac{\ell+L+4}{2}) } 
  \equiv \beta_{\ell L}^{T, \rm SW}  ~.
\end{eqnarray}
Substituting these and a further small-$\ell$ approximation: $\omega_{\ell L}(x_{\rm ls}) \approx -5 \beta_{\ell L}^{T, \rm SW}$, into Eq.~\eqref{eq:blllTTmu_all_dL}, we obtain the SW-limit formula:
\begin{eqnarray}
  b_{\ell_1 \ell_2 \ell_3, \rm SW}^{TT\mu, L = \rm even}
  &=& 9 d_L A_S \ln\left(\frac{k_i}{k_f}\right)
\frac{4\pi}{2L+1}  \sum_{L_1 L_3}
(-1)^{\frac{L_1 + L_3 + \ell_1 + \ell_3}{2}}
\frac{h_{L_1 \ell_2 L_3}}{h_{\ell_1 \ell_2 \ell_3}}
h_{\ell_1 L_1 L}h_{\ell_3 L_3 L}
  \left\{
  \begin{array}{ccc}
  \ell_1 & \ell_2 & \ell_3 \\
  L_3 & L & L_1 
  \end{array}
  \right\} 
 \nonumber \\ 
 &&  \times \beta_{\ell_1 L_1}^{T, \rm SW} \beta_{\ell_3 L_3}^{T, \rm SW}
 \times
 \begin{cases}
   3 &: L = 0  \\
   1 &: L \geq 2
    \end{cases}
\, + \, (\ell_1 \leftrightarrow \ell_2) ~. \label{eq:blllTTmu_all_dL_SW}
  \end{eqnarray}

Figure~\ref{fig:blllTTmu_3D} shows the shape difference between $b_{\ell_1 \ell_2 \ell_3}^{TT\mu, d_2}$, $b_{\ell_1 \ell_2 \ell_3}^{TT\mu, d_0}$ (or equivalently $TT\mu$ from the $\tau_{\rm NL}$-type trispectrum, $b_{\ell_1 \ell_2 \ell_3}^{TT\mu, \tau_{\rm NL}}$ \cite{Bartolo:2015fqz}), and $TT\mu$ from the $g_{\rm NL}$-type trispectrum, $b_{\ell_1 \ell_2 \ell_3}^{TT\mu, g_{\rm NL}}$ \cite{Bartolo:2015fqz}. We can see from the left panels that both $b_{\ell_1 \ell_2 \ell_3}^{TT\mu, d_0}$ and $b_{\ell_1 \ell_2 \ell_3}^{TT\mu, d_2}$ are both mostly enhanced at $\ell_1 \sim \ell_2 \gg \ell_3$. However their color patterns are not the same, due to the different angle dependence, as we have seen in Eq.~\eqref{eq:ratio_blllTTmu_L0_L2}. These color distributions are also quite different from that of $b_{\ell_1 \ell_2 \ell_3}^{TT\mu, g_{\rm NL}}$. This visual inspection prompts the expectation that the $TT\mu$ bispectra generated by the different types of trispectra we are considering will display a very low level of correlation, as indeed confirmed in Fig.~\ref{fig:corr_coeff_lmu1000}, where correlation coefficients are explicitly computed. These shapes can thus be clearly distinguished using $TT \mu$.

Both the exact \eqref{eq:blllTTmu_all_dL} and the SW-limit \eqref{eq:blllTTmu_all_dL_SW} results will be employed in the Fisher matrix computations, discussed in the next section.

\section{Forecasts}\label{sec:fisher}

\begin{figure}[t!]
  \begin{tabular}{c}
    \begin{minipage}{1.0\hsize}
      \begin{center}
    \includegraphics[width=0.8\textwidth]{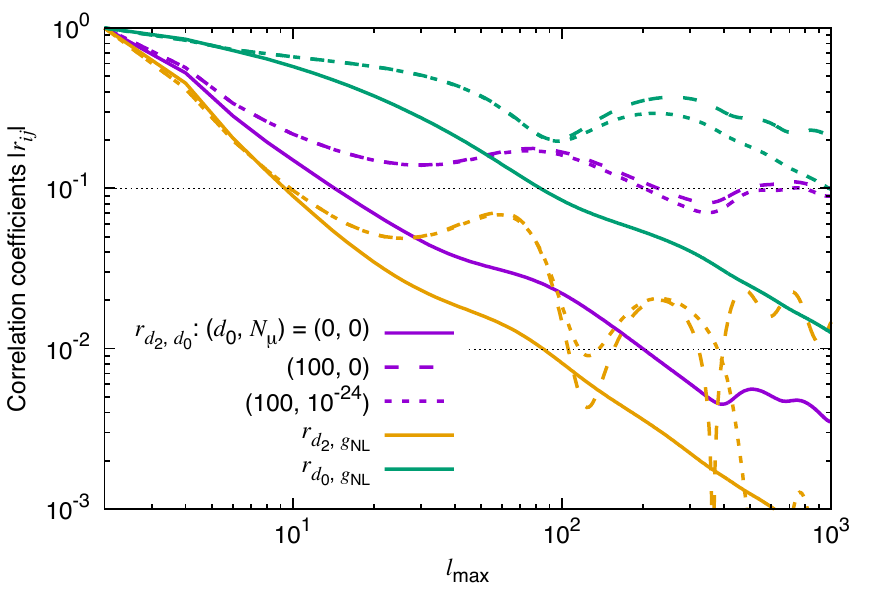}
      \end{center}
    \end{minipage}
  \end{tabular}
  \caption{Correlation coefficients between $b_{\ell_1 \ell_2 \ell_3}^{TT\mu, d_2}$, $b_{\ell_1 \ell_2 \ell_3}^{TT\mu, d_0}$ and $b_{\ell_1 \ell_2 \ell_3}^{TT\mu, g_{\rm NL}}$ for $(d_0, N_\mu) = (0,0)$ (solid lines), $(100, 0)$ (dashed lines) and $(100, 10^{-24})$ (dotted lines) with $\ell_{\mu} = 1000$ as a function of $\ell_{\rm max}$. The results for $(d_0, N_\mu) = (0, 10^{-24})$ (undescribed here) overlap substantially with the solid lines. Higher values tend to be induced by nonzero $C_\ell^{\mu\mu, d_0}$.} \label{fig:corr_coeff_lmu1000}
\end{figure}

In this section we discuss the detectability of the angle-dependent curvature trispectrum \eqref{eq:zeta4_def} in future surveys. We focus especially on $d_2$, since it is the lowest order mode (physically motivated) characterizing a non-trivial angular dependence and producing distinctive features in $TT\mu$. In fact a non-zero value of $d_2$ is predicted in concrete inflationary models such as the vector field model, along with $d_0 (= \tau_{\rm NL} / 6)$ \cite{Shiraishi:2013vja,Abolhasani:2013zya,Shiraishi:2013oqa,Rodriguez:2013cj}.

Let us consider the Fisher matrix:
\begin{eqnarray}
 F_{ij} = \sum_{\ell_1, \ell_2, \ell_3 = 2}^{\ell_{\rm max}} h_{\ell_1 \ell_2 \ell_3}^2  \frac{\hat{b}_{\ell_1 \ell_2 \ell_3}^{TT\mu, i} \hat{b}_{\ell_1 \ell_2 \ell_3}^{TT\mu, j}}{2 C_{\ell_1}^{TT} C_{\ell_2}^{TT} C_{\ell_3}^{\mu\mu}} ~, \label{eq:fish_TTmu}
\end{eqnarray}
where $\hat{b}_{\ell_1 \ell_2 \ell_3}^{TT\mu}$ is the $TT\mu$ bispectrum normalized to the parameter under examination (i.e. either $b_{\ell_1 \ell_2 \ell_3}^{TT\mu, d_L = 1}$ or $b_{\ell_1 \ell_2 \ell_3}^{TT\mu, g_{\rm NL} = 1}$). This expression for the Fisher matrix is valid under assumptions that off-diagonal components, all NG contributions and the $T\mu$ correlation are negligible in the covariance matrix; the latter condition can be expressed as $(C_\ell^{T\mu})^2 \ll C_\ell^{TT} C_{\ell}^{\mu\mu}$. The observed $\mu\mu$ power spectrum is given by the sum of the signal from the Gaussian part of curvature perturbations, the isotropic NG part from $d_0$, and the instrumental noise spectrum, reading $C_\ell^{\mu\mu} = C_\ell^{\mu\mu, \rm G} + C_\ell^{\mu\mu, d_0} + N_\mu e^{\ell^2 / \ell_\mu^2}$, with $C_\ell^{\mu\mu, \rm G} \sim 10^{-30}$ and $C_\ell^{\mu\mu, d_0} \simeq 3.3 \times 10^{-22} d_0 \ell^{-2}$ (Eq.~\eqref{eq:Clmumu_d0} or \cite{Pajer:2012vz,Bartolo:2015fqz}). Note that the signal from $d_{L \geq 1}$ and $g_{\rm NL}$ are not included in the above $C_\ell^{\mu\mu}$ because they are subdominant (see Appendix~\ref{appen:power} and \cite{Bartolo:2015fqz}). Uncertainties due to instrumental noise may be ignored in $C_\ell^{TT}$, since CMB temperature fluctuations have already been measured with close to CVL-level accuracy, for the $\ell$-range under exam.

It has been visually confirmed from Fig.~\ref{fig:blllTTmu_3D} that the shapes of $b_{\ell_1 \ell_2 \ell_3}^{TT\mu, d_0}$, $b_{\ell_1 \ell_2 \ell_3}^{TT\mu, d_2}$ and $b_{\ell_1 \ell_2 \ell_3}^{TT\mu, g_{\rm NL}}$ look very different. This can be expressed in a more quantitative way by computing the correlation coefficients: $r_{ij} \equiv F_{ij} / \sqrt{ F_{ii} F_{jj}}$. Numerical results for several $d_0$ and $N_\mu$ with $\ell_\mu = 1000$, summarized in Fig.~\ref{fig:corr_coeff_lmu1000}, lead us to conclude that $|r_{d_2, d_0}| \lesssim 0.1$, $|r_{d_2, g_{\rm NL}}| \lesssim 0.01$ and $|r_{d_0, g_{\rm NL}}| \lesssim 0.2$ at $\ell_{\rm max} = 1000$. The parameter $d_0$, $d_2$ and $g_{\rm NL}$, estimated via $TT \mu$, are thus close to be uncorrelated. The expected $1\sigma$ error $\Delta_i$, on the $i$-th parameter can therefore be computed directly from the corresponding diagonal elements of the Fisher matrix, as $\Delta_i = 1 / \sqrt{F_{ii}}$.

\subsection{Detectability of $d_2$}

\begin{figure}[t!]
  \begin{tabular}{c}
    \begin{minipage}{1.0\hsize}
  \begin{center}
    \includegraphics[width=0.8\textwidth]{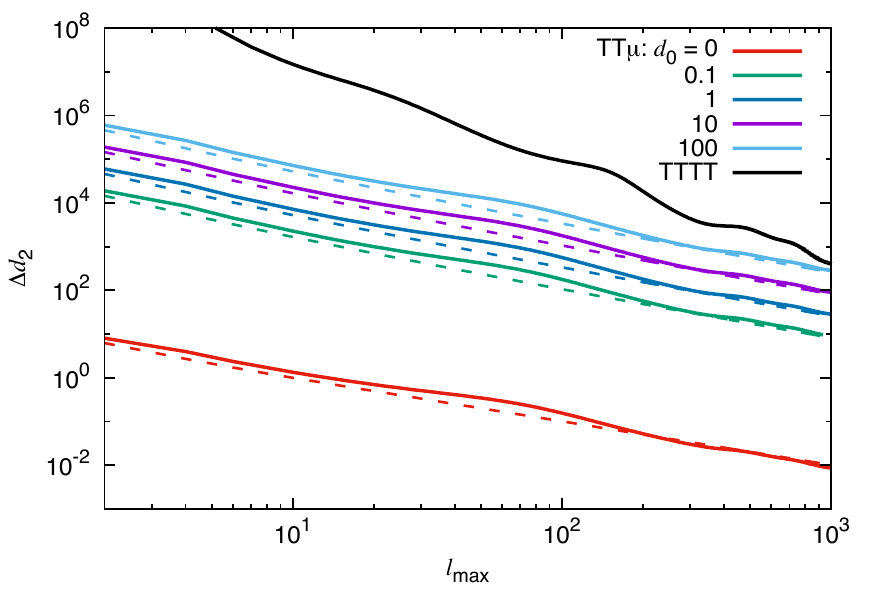}
  \end{center} 
    \end{minipage}
  \end{tabular}
  \caption{Expected $1\sigma$ errors on $d_2$ in a noiseless CVL-level measurement of $TT\mu$ with $N_\mu = 0$, for several nonzero $d_0$. 
    Solid and dashed lines are obtained using the full-sky $TT\mu$ expression \eqref{eq:blllTTmu_all_dL} and the SW approximation \eqref{eq:blllTTmu_all_dL_SW}, respectively. The black line describes $\Delta d_2$ from the CMB temperature trispectrum ($TTTT$), computed in \cite{Shiraishi:2013oqa}.} \label{fig:dd2_CV}
\end{figure}

\begin{figure}[t!]
  \begin{tabular}{c}
    \begin{minipage}{1.0\hsize}
  \begin{center}
    \includegraphics[width=0.8\textwidth]{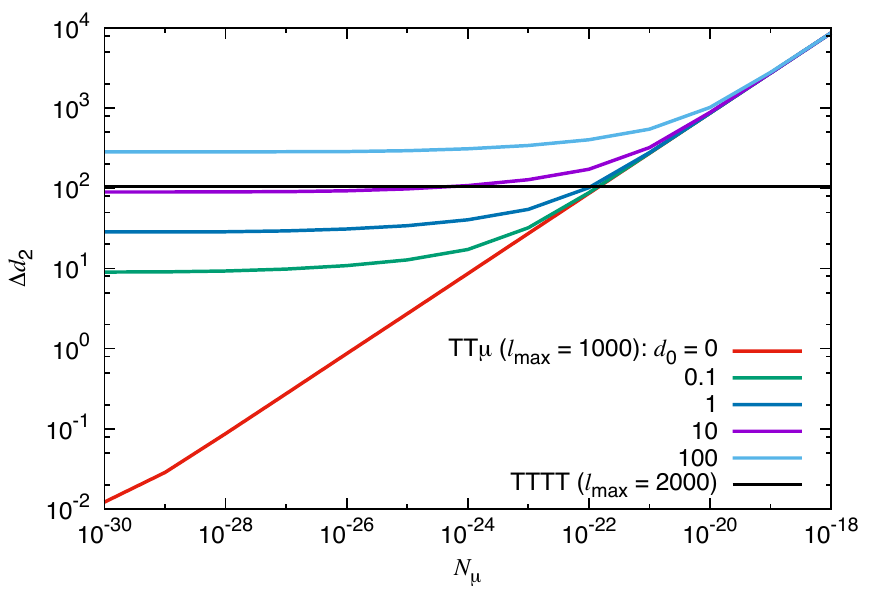}
  \end{center}
    \end{minipage}
  \end{tabular}
   \caption{Expected $1\sigma$ errors on $d_2$ obtained from $TT\mu$ measurements with $\ell_{\rm max} = 1000$, as a function of $N_\mu$ with $\ell_\mu = 1000$. For comparison, we also plot $\Delta d_2$ from $TTTT$ with $\ell_{\rm max} = 2000$ \cite{Shiraishi:2013oqa}.} \label{fig:Nmu_vs_dd2_lmu1000}
\end{figure}

Figure~\ref{fig:dd2_CV} describes $1\sigma$ errors on $d_2$ ($\Delta d_2$) estimated from the Fisher matrix for $TT\mu$ \eqref{eq:fish_TTmu} in a CVL measurement with $ N_\mu = 0$. Since $b_{\ell_1 \ell_2 \ell_3}^{TT\mu, d_2}$ is enhanced in the squeezed limit: $\ell_1 \sim \ell_2 \gg \ell_3$, like $b_{\ell_1 \ell_2 \ell_3}^{TT\mu, d_0}$ or $b_{\ell_1 \ell_2 \ell_3}^{TT\mu, \tau_{\rm NL}}$ (see Fig.~\ref{fig:blllTTmu_3D}), $\Delta d_2$ scales in the same way as $\Delta \tau_{\rm NL}$, as $\ell_{\rm max}^{-1}$ \cite{Bartolo:2015fqz}. In contrast, as described in Fig.~\ref{fig:dd2_CV}, $\Delta d_2$ estimated from the CMB temperature trispectrum ($TTTT$) scales more rapidly, like $\ell_{\rm max}^{-2}$ \cite{Shiraishi:2013oqa}. Despite this disadvantage, interestingly, $TT\mu$ can outperform $TTTT$ for $\ell_{\rm max} < 1000$ even if $d_0$ increases up to $\sim 100$. For a perfectly ideal scenario, with $d_0 = 0$ and $N_\mu = 0$, the smallest detectable value of $d_2$ is $d_2 \sim 0.01$.

On the other hand, in realistic experiments such as {\it Planck} \cite{Planck:2006aa}, PIXIE \cite{Kogut:2011xw} and CMBpol \cite{Baumann:2008aq}, nonzero instrumental noise and finite angular resolution will worsen sensitivities. Including the following specifications: $(N_\mu, \ell_\mu) = (10^{-15}, 861)$ ({\it Planck}), $(10^{-17}, 84)$ (PIXIE) and $(2\times 10^{-18}, 1000)$ (CMBpol) \cite{Ganc:2012ae,Ganc:2014wia}, into the Fisher matrix~\eqref{eq:fish_TTmu}, we obtain  $\Delta d_2^{TT\mu}|_{ d_0 = 0} = 2.7 \times 10^5$ ({\it Planck}), $2.7 \times 10^4$ (PIXIE) and $1.2 \times 10^4$ (CMBpol) at $\ell_{\rm max} = 1000$.

In Fig.~\ref{fig:Nmu_vs_dd2_lmu1000}, with more futuristic $TT\mu$ measurements in mind, we extend $N_\mu$ from $10^{-18}$ to $10^{-30}$. One can see that the expected $d_2$-sensitivity plateaus when $N_\mu$ becomes very small and cosmic variance ($C_\ell^{\mu\mu, \rm G} + C_\ell^{\mu\mu, d_0}$) dominates over $C_\ell^{\mu\mu}$. Therefore, when $d_0 \gtrsim 10$, $\Delta d_2^{TT\mu}$ becomes flat so quickly (due to the large value of $C_\ell^{\mu\mu, d_0}$) that $TT\mu$ cannot outperform $TTTT$, even for very small $\mu$ noise levels, $N_\mu$. In order to get $\Delta d_2^{TT\mu} < \Delta d_2^{TTTT}$, $d_0 \lesssim 1$ and $N_\mu \lesssim 10^{-22}$ are required. The ratio of $\Delta d_2^{TT\mu}$ to $\Delta d_0^{TT\mu}$ is nearly constant with $d_0$ or $N_\mu$. Our result, $\Delta d_2^{TT\mu} / \Delta d_0^{TT\mu} \simeq 11$, is larger than $\Delta d_2^{TTTT} / \Delta d_0^{TTTT} \simeq 4$ \cite{Shiraishi:2013oqa}.

\subsection{Detectability of $g_*$ in the $f(\phi)F^2$ model}

\begin{figure}[t!]
  \begin{tabular}{c}
    \begin{minipage}{1.0\hsize}
      \begin{center}
    \includegraphics[width=0.8\textwidth]{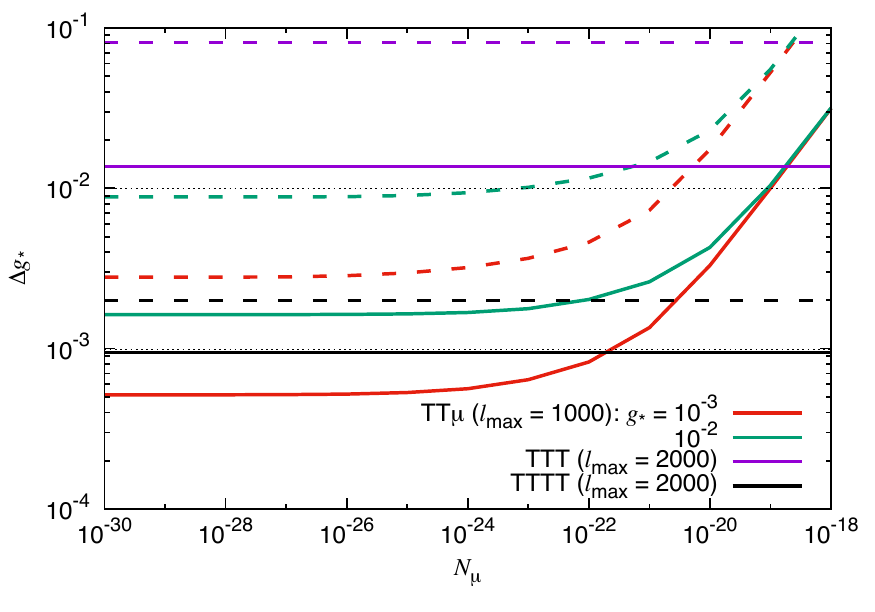}
  \end{center}
    \end{minipage}
    \end{tabular}
\\
\begin{tabular}{c}
    \begin{minipage}{1.0\hsize}
  \begin{center}
    \includegraphics[width=0.8\textwidth]{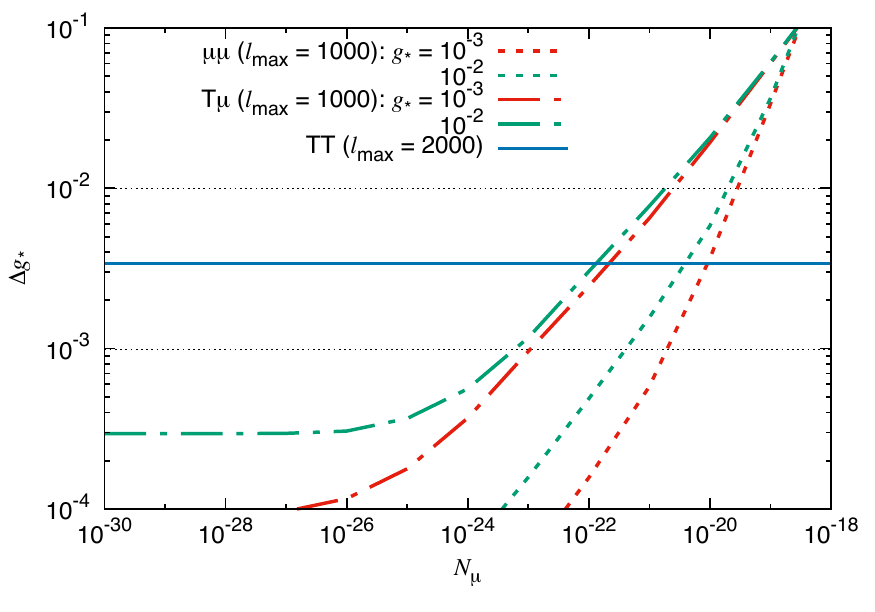}
  \end{center}
\end{minipage}
  \end{tabular}
  \caption{Expected errors on $g_*$ in the $f(\phi)F^2$ model with $f(\phi) \propto a^{-4}$ and $N = 60$, translated from $\Delta d_{0,2}^{TT\mu}$, $\Delta d_{0,2}^{TTTT}$, $\Delta c_{0,2}^{TTT}$, $\Delta d_0^{\mu\mu}$ and $\Delta c_0^{T\mu}$, and $\Delta g_*$ obtained from $TT$. The results from $TT\mu$, $\mu\mu$ and $T\mu$ are computed in terms of $g_* = 10^{-2}$ and $ 10^{-3}$ and $10^{-30}\leq N_\mu \leq 10^{-18}$, with $\ell_\mu = 1000$. The solid (dashed) lines in the top panel correspond to $\Delta g_*$ translated from $\Delta c_0$ or $\Delta d_0$ ($\Delta c_2$ or $\Delta d_2$).} \label{fig:Nmu_vs_dgstar_fFF_N60_lmu1000}
\end{figure}

In the above analysis, we treat $d_0$ and $d_2$ as independent parameters. However, upon specifying the inflationary model, these can be related to each other.

For a practical example, here, let us consider an inflationary model where the inflaton field couples to a U(1) gauge field with a non-vanishing vacuum expectation value via a $f(\phi) F^2$ interaction. The directional dependence of the gauge field $\hat{A}$ is now directly imprinted on the correlators of curvature perturbations via such interaction. The time dependence of $f(\phi)$ controls the scale dependence of the correlation functions, and a nearly scale-invariant shape can be realized by choosing $f(\phi) \propto a^{-4}$, with $a(\tau)$ denoting the scale factor. Since effects of the gauge field on the curvature perturbation, via $f(\phi) F^2$, are always quadratic, a quadrupolar modulation, $g_* (\hat{k} \cdot \hat{A})^2$, is generated in the power spectrum \cite{Ackerman:2007nb,Soda:2012zm,Naruko:2014bxa}. Moreover, nonzero $c_2$ and $d_2$ components arise in the bispectrum \cite{Bartolo:2012sd,Shiraishi:2013vja} and the trispectrum \cite{Shiraishi:2013vja,Shiraishi:2013oqa,Abolhasani:2013zya,Rodriguez:2013cj}, in addition to $c_0$ and $d_0$. These parameters are then related to each other and we can express $c_L$ and $d_L$ by means of a single parameter $g_*$ as \cite{Shiraishi:2013vja,Shiraishi:2013oqa}
\begin{eqnarray}
  c_0 &=& 2 c_2 \approx 3.2 \frac{|g_*|}{0.01} \frac{N}{60} ~, \ \
  c_1 = c_3 = c_4 = \cdots = 0 ~, \label{eq:cL_gstar_fFF} \\
  d_0 &=& \frac{d_2}{2} \approx 2.6 \times 10^2 \frac{|g_*|}{0.01} \left(\frac{N}{60}\right)^2 ~, \ \
d_1 = d_3 = d_4 = \cdots = 0 ~, 
  \label{eq:dL_gstar_fFF}
\end{eqnarray} 
where $N$ is the number of e-folds, before the end of inflation, at which the observed modes leave the horizon. Note that the approximations in these equations are reasonable, especially for large-field inflation models \cite{Naruko:2014bxa}. These relations lead to $(C_\ell^{T\mu})^2 \ll C_\ell^{TT} C_\ell^{\mu\mu}$, allowing us to use the Fisher matrix forms~\eqref{eq:fish_TTmu}, \eqref{eq:fish_Tmu} and \eqref{eq:fish_mumu} in our forecasts (See Appendix~\ref{appen:power} for details).

In Fig.~\ref{fig:Nmu_vs_dgstar_fFF_N60_lmu1000}, we plot $1\sigma$ errors on $g_*$ ($\Delta g_*$), translated via a Fisher matrix analysis from $\Delta d_{0,2}^{TT\mu}$, $\Delta d_{0,2}^{TTTT}$, $\Delta c_{0,2}^{TTT}$, $\Delta d_0^{\mu\mu}$ and $\Delta c_0^{T\mu}$, via Eqs.~\eqref{eq:cL_gstar_fFF} and \eqref{eq:dL_gstar_fFF}, and $\Delta g_*$ from $TT$. Note that, as $\mu\mu$ and $T\mu$ vanish, except for $L = 0$ (see Appendix~\ref{appen:power} for details), we plot neither $\Delta g_*$ from $\Delta d_2^{\mu\mu}$ nor that from $\Delta c_2^{T\mu}$. For $TT\mu$, $\mu\mu$ and $T\mu$, we examine the cases with $g_* = 10^{-2}$ and $10^{-3}$ (since, e.g, it also enters into $C_\ell^{\mu\mu}$ in Eq.~\eqref{eq:fish_TTmu}) and vary $N_\mu$ from $10^{-30}$ to $10^{-18}$ with $\ell_\mu = 1000$ in the same manner as Fig.~\ref{fig:Nmu_vs_dd2_lmu1000}. The similar $N_\mu$ dependence as $\Delta d_2^{TT\mu}$; namely, $\Delta g_* \propto \sqrt{N_\mu}$ for large $N_\mu$ and $\Delta g_* = {\rm const.}$ for small $N_\mu$, is confirmed from the $TT\mu$ lines in the top panel, as expected. Moreover, we notice that $\Delta g_*$ from $TT\mu$ is boosted by the increase of $g_*$. This leads to the result that, in $d_2$ measurements based on $TT\mu$, $g_* = 10^{-3}$ is, in principle, undetectable (i.e., $\Delta g_* > g_*$ with any small $N_\mu$). However, $g_* = 10^{-2}$, comparable to the latest upper bound from $TT$ \cite{Kim:2013gka,Ade:2015sjc,Ade:2015lrj} and to the smallest detectable value from $TTT$ \cite{Shiraishi:2013vja,Shiraishi:2013oqa}, is measurable if $N_\mu \lesssim 10^{-24}$. If we consider the $d_0$ component, $TT\mu$ achieves better sensitivity and $g_* = 10^{-3}$, comparable to the smallest detectable value from $TTTT$ \cite{Shiraishi:2013oqa}, is observable if we reduce $N_\mu$ below $\sim 10^{-22}$. On the other hand, we realize from the bottom panel in Fig.~\ref{fig:Nmu_vs_dgstar_fFF_N60_lmu1000} that the best limits on $\Delta g_*$ can be obtained from $\mu\mu$ and $T\mu$. In this sense, $TT\mu$ will be useful to cross-check a possible nonzero $g_*$ signal observed in $\mu\mu$ and $T\mu$. 

\section{Conclusions}\label{sec:conclusions}

If some anisotropic source is present in the very Early Universe, the primordial trispectra of curvature perturbations can display a characteristic, non-trivial angular dependence between different wavenumbers ${\bf k}$, which can be expressed in terms of Legendre polynomials. This paper discussed the possibility to observe such angular dependence using a new type of observable, recently found in \cite{Bartolo:2015fqz}, namely the $TT\mu$ correlation function generated from CMB temperature and $\mu$-distortion anisotropies. For the sake of intuitive understanding, we started our calculation of angular-dependent $TT \mu$ by employing the flat-sky approximation (in analogy with the previous CMB temperature trispectrum analysis done in \cite{Shiraishi:2013oqa}), and verified that the specific angular dependence in ${\bf k}$--space gets directly projected to $\boldsymbol{\ell}$-space. Therefore, $TT\mu$ changes its amplitude and sign, depending on the angle between each $\boldsymbol{\ell}$.

After this preliminary calculation, we performed a more accurate full sky quantitative analysis. Using a Fisher matrix approach, we found that $TT\mu$ from the $L=2$ mode in the Legendre-type template \eqref{eq:zeta4_def} is nearly orthogonal to $TT\mu$ from the $L=0$ mode (or equivalently the $\tau_{\rm NL}$-type trispectrum) and $TT\mu$ from the $g_{\rm NL}$-type trispectrum. This is an important feature when it comes to discriminating between shapes. Our parameter forecasts showed that, in the absence of the $L=0$ mode (i.e., $d_0 = 0$), a CVL-level measurement of $\mu$-distortion fluctuations enables us to detect the $L=2$ mode with $d_2 \sim 0.01$ sensitivity, which is $4$ orders of magnitude smaller than the value accessible by the temperature trispectrum ($TTTT$). Even in more realistic cases, $TT\mu$ could outperform $TTTT$, although instrumental uncertainties and additional cosmic variance, generated by nonzero $d_0$, reduce the sensitivity to $d_2$. Once fixing the inflationary model, the parameters of the power spectrum, bispectrum  and trispectrum are related to each other. Considering the $f(\phi)F^2$ model, and employing the consistency relation \eqref{eq:dL_gstar_fFF}, we reach the conclusion that a quadrupolar power asymmetry with $g_* \sim 10^{-3}$ could, in principle, be detected from $TT\mu$.


\acknowledgments
We thank James R. Fergusson for helping to draw beautiful 3D representations of the $TT\mu$ bispectra. MS was supported in part by a Grant-in-Aid for JSPS Research under Grants No.~27-10917, and in part by the World Premier International Research Center Initiative (WPI Initiative), MEXT, Japan. This work was supported in part by ASI/INAF Agreement I/072/09/0 for the Planck LFI Activity of Phase E2. Numerical computations were in part carried out on Cray XC30 at Center for Computational Astrophysics, National Astronomical Observatory of Japan.


\appendix

\section{The $T\mu$ and $\mu\mu$ power spectrum}\label{appen:power}

We here examine the angular power spectra of $T\mu$ from the angle-dependent bispectrum \eqref{eq:zeta3_def} and $\mu\mu$ from the angle-dependent trispectrum \eqref{eq:zeta4_def}. These are used to compute the error bar $\Delta g_*$ within the $f(\phi)F^2$ model of Sec.~\ref{sec:fisher}. 

With Eqs.~\eqref{eq:alm_mu_full} and \eqref{eq:alm_T_full}, the $T\mu$ correlation is formulated as
\begin{eqnarray}
 \Braket{a_{\ell_1 m_1}^T a_{\ell_2 m_2}^\mu }
  &=& 4\pi i^{\ell_1} \int \frac{d^3 {\bf k}_1}{(2\pi)^{3}}
  {\cal T}_{\ell_1}(k_1)  Y_{\ell_1 m_1}^*(\hat{k}_1) 
  4\pi (-i)^{\ell_2} \left[ \prod_{n=1}^2 
\int \frac{d^3 {\bf K}_n}{(2\pi)^3} 
\right] \int d^3 {\bf K}_3 \nonumber \\ 
&& \times \delta^{(3)}\left(\sum_{n=1}^3 {\bf K}_n \right) 
Y_{\ell_2 m_2}^*(\hat{K}_3)
j_{\ell_2}(K_3 x_{\rm ls}) 
f(K_1, K_2, K_3) 
\Braket{\zeta_{{\bf k}_1} \zeta_{{\bf K}_1} \zeta_{{\bf K}_2}} ~.
\end{eqnarray}
Plugging Eq.~\eqref{eq:zeta3_def} into this and evaluating with the squeezed-limit filtering by $f(K_1, K_2, K_3)$ yield 
\begin{eqnarray}
   \Braket{a_{\ell_1 m_1}^T a_{\ell_2 m_2}^\mu }
  &\simeq& i^{\ell_1- \ell_2}  \int \frac{d^3 {\bf k}_1}{2\pi^2}
  {\cal T}_{\ell_1}(k_1)  Y_{\ell_1 m_1}^*(\hat{k}_1) 
Y_{\ell_2 m_2}^*(\hat{k}_1) 
j_{\ell_2}(k_1 x_{\rm ls}) \int_0^\infty \frac{K_1^2 d K_1}{2\pi^2} f(K_1, K_1, k_1)
 \nonumber \\
&&  \times  
\int d^2 \hat{K}_1
\sum_L c_L \left[ 1 + (-1)^{L} \right]
    {\cal P}_L(\hat{k}_1 \cdot \hat{K}_1) P(k_1) P(K_1)
~.
\end{eqnarray}
We notice that the contribution of the $L \geq 1$ mode vanishes since $\int d^2 \hat{K}_1 {\cal P}_{L \geq 1}(\hat{k}_1 \cdot \hat{K}_1) = 0$. We therefore obtain $\Braket{a_{\ell_1 m_1}^T a_{\ell_2 m_2}^\mu} = C_{\ell_1}^{T\mu}(-1)^{m_1} \delta_{\ell_1, \ell_2} \delta_{m_1, -m_2}$ with 
\begin{eqnarray}
  C_{\ell}^{T\mu} \simeq  C_\ell^{T\mu, c_0} 
  = 18 \pi c_0 A_S^2 \ln\left(\frac{k_i}{k_f}\right) 
  \int_0^\infty \frac{d k}{k} 
     {\cal T}_{\ell}(k)  j_{\ell}(k x_{\rm ls})
     e^{-k^2 / k_{\rm rec}^2} ~. \label{eq:ClTmu_c0} 
\end{eqnarray}
In the same manner, one can verify that $\mu\mu$ from the $L \geq 1$ mode of Eq.~\eqref{eq:zeta4_def} is highly suppressed. The angular power spectrum reads $\Braket{a_{\ell_1 m_1}^\mu a_{\ell_2 m_2}^\mu} = C_{\ell_1}^{\mu\mu}(-1)^{m_1} \delta_{\ell_1, \ell_2} \delta_{m_1, -m_2}$ with
\begin{eqnarray}
C_{\ell}^{\mu\mu} \simeq C_{\ell}^{\mu \mu , d_0} 
= 243 \pi d_0
 A_S^3
\left[ \ln\left( \frac{k_i}{k_f} \right) \right]^2
\frac{1}{\ell(\ell+1)} ~. \label{eq:Clmumu_d0}
\end{eqnarray}

The expected errors on $c_0$ and $d_0$, described in the bottom panel of Fig.~\ref{fig:Nmu_vs_dgstar_fFF_N60_lmu1000}, are computed, respectively, as $\Delta c_0^{T\mu} = 1/\sqrt{F^{T\mu}}$ and $\Delta d_0^{\mu\mu} = 1/\sqrt{F^{\mu\mu}}$ with 
\begin{eqnarray}
  F^{T\mu} &=& \sum_{\ell=2}^{\ell_{\rm max}} (2\ell + 1) \frac{\left(C_\ell^{T\mu, c_0 = 1}\right)^2}{C_\ell^{TT} C_\ell^{\mu\mu}} ~, \label{eq:fish_Tmu} \\
  F^{\mu\mu} &=& \sum_{\ell=2}^{\ell_{\rm max}} \frac{2\ell + 1}{2} \left(\frac{C_\ell^{\mu\mu, d_0 = 1}}{C_\ell^{\mu\mu}}\right)^2~. \label{eq:fish_mumu}
  \end{eqnarray}
These indicate $\Delta c_0^{T\mu} \propto \sqrt{N_\mu}$ and $\Delta d_0^{\mu\mu} \propto N_\mu$ for large $N_\mu$, agreeing with numerical results in the bottom panel of Fig.~\ref{fig:Nmu_vs_dgstar_fFF_N60_lmu1000}. This difference realizes the outperformance of $\mu\mu$ for small $N_\mu$.

Considering the consistency relations in the $f(\phi)F^2$ model \eqref{eq:cL_gstar_fFF} and \eqref{eq:dL_gstar_fFF} and the SW approximation for $T\mu$, we obtain
\begin{eqnarray}
  C_{\ell, \rm SW}^{T\mu} &\simeq& - 5.9 \times 10^{-16}
  \frac{|g_*|}{0.01} \frac{N}{60}  \frac{1}{\ell(\ell+1)}
  ~, \\
  C_{\ell}^{\mu\mu} &\simeq&  8.4 \times 10^{-20} \frac{|g_*|}{0.01} \left(\frac{N}{60}\right)^2  \frac{1}{\ell (\ell+1)} ~.
\end{eqnarray}
These and $C_{\ell, \rm SW}^{TT} \simeq 6.0 \times 10^{-10} / [\ell (\ell+1) ]$ result in 
\begin{eqnarray}
  \frac{\left(C_{\ell, \rm SW}^{T\mu}\right)^2}{C_{\ell, \rm SW}^{TT} C_{\ell}^{\mu\mu}}
  \simeq 6.9 \times 10^{-3}  \frac{|g_*|}{0.01} ~,
  \end{eqnarray}
indicating that $C_\ell^{T\mu}$ can be ignored in the Fisher matrix for $g_*$ in the $f(\phi)F^2$ model (because of the observational constraints $|g_*| \lesssim 0.01$ \cite{Kim:2013gka,Ade:2015lrj,Ade:2015sjc}).

\section{Constant model} \label{appen:constmodel}

We here derive $TT\mu$ sourced from the constant curvature trispectrum:
\begin{equation}
  \Braket{\prod_{n=1}^4 \zeta_{{\bf k}_n} }
  = (2\pi)^3 \delta^{(3)}\left(\sum_{n=1}^4 {\bf k}_n  \right)
  \left[ P(k_1)P(k_2)P(k_3)P(k_4) \right]^{3/4} ~,
\end{equation}
used as a normalization in Fig.~\ref{fig:blllTTmu_3D}. The normalization by the constant model template has been originally employed to draw the CMB bispectra \cite{Fergusson:2009nv,Fergusson:2010dm} and trispectra \cite{Regan:2010cn}. 

This trispectrum is independent of wavenumbers of the sum of two wavevectors, such as $|{\bf k}_1 + { \bf k}_2|$, and hence has a similar structure to the $g_{\rm NL}$-type trispectrum. By the application of the approach for the $g_{\rm NL}$ case \cite{Bartolo:2015fqz}, we can obtain a form reasonable for $\ell_3 \leq k_{\rm rec} x_{\rm ls} \sim 2000$:
\begin{equation}
  b_{\ell_1 \ell_2 \ell_3}^{TT\mu, \rm const}
 \simeq 
  \int_0^\infty r^2 dr 
  \eta_{\ell_1}^T(r) \eta_{\ell_2}^T(r)
\sum_{L_1 L_2} \frac{h_{L_1 L_2 \ell_3}^2}{2\ell_3 + 1}
  \left[ \eta_{L_1}^\mu(r,z_i) \eta_{L_2}^\mu(r,z_i) - \eta_{L_1}^\mu(r,z_f) \eta_{L_2}^\mu(r,z_f)\right],
\end{equation}
where 
\begin{eqnarray}
  \eta_\ell^T(r) &\equiv&  \frac{2}{\pi} \int_0^\infty k^2 d k P^{3/4}(k) {\cal T}_{\ell}(k) j_{\ell}(k r) ~, \\
  \eta_L^\mu(r,z) &\equiv& \frac{3}{\pi} \int_0^\infty k^2 dk P^{3/4}(k) 
  j_{L}(k x_{\rm ls}) j_{L}(k r) e^{-k^2/k_D^2(z)} ~.
  \end{eqnarray}
Both $\eta_{\ell}^T(r)$ and $\eta_L^\mu(r,z)$ have peaks at $r \sim x_{\rm ls}$. Owing to a fact that $\eta_L^\mu(x_{\rm ls},z)$ depends weakly on $k_D(z)$ for $L \lesssim k_D(z) x_{\rm ls}$ and decays rapidly for $L \gtrsim k_D(z) x_{\rm ls}$ (like $\beta_L^\mu(x_{\rm ls},z)$ in \cite{Bartolo:2015fqz}), and the triangular inequality of $h_{L_1 L_2 \ell_3}$, we are allowed to evaluate with $\sum_{L_1 L_2} h_{L_1 L_2 \ell_3}^2 \simeq (2\ell_3 + 1) \sum_{L_1,L_2 = L_f}^{L_i} (2 L_1 / \pi^2) \delta_{L_1, L_2}$, where $ L_f \equiv k_f x_{\rm ls} \sim 10^5$ and $ L_i \equiv k_i x_{\rm ls} \sim 10^8$, and finally reach
\begin{eqnarray}
  b_{\ell_1 \ell_2 \ell_3}^{TT\mu, \rm const}
 \simeq  
  \int_0^\infty r^2 dr 
 \eta_{\ell_1}^T(r) \eta_{\ell_2}^T(r) 
 \sum_{L_1 = L_f}^{L_i}  \frac{2 L_1}{ \pi^2}
 \left[ \eta_{L_1}^\mu(r,z_i) \right]^2
    ~.
\end{eqnarray}
Since $\ell_1, \ell_2 \ll L_f < L_1 < L_i $, $\eta_{L_1}^{\mu}(r,z_i)$ is sharply peaked at $r \sim x_{\rm ls}$, compared with $\eta_{\ell_1}^{T}(r)$ and $\eta_{\ell_2}^{T}(r)$. Owing to this, the interval of the $r$ integral is practically limited to a very narrow window by $\eta_{L_1}^{\mu}(r,z_i)$, and $\eta_{\ell_1}^{T}(r)$ and $\eta_{\ell_2}^{T}(r)$ remain almost unchanged there. This enables us to move $\eta_{\ell_1}^{T}(r)$ and $\eta_{\ell_2}^{T}(r)$ outside the $r$ integral with the evaluation at $r = x_{\rm ls}$ as  
\begin{eqnarray}
  b_{\ell_1 \ell_2 \ell_3}^{TT\mu, \rm const}
  &\simeq&
  \eta_{\ell_1}^T(x_{\rm ls}) \eta_{\ell_2}^T(x_{\rm ls})
  \int_0^\infty r^2 dr 
   \sum_{L_1 = L_f}^{L_i}  \frac{2 L_1}{ \pi^2}
 \left[ \eta_{L_1}^\mu(r,z_i) \right]^2
 ~.
\end{eqnarray}

The $r$ integral and the $L_1$ summation now give just a dimensional number, so $\eta_{\ell_1}^T(x_{\rm ls})$ and $\eta_{\ell_2}^T(x_{\rm ls})$ are responsible for the $\ell$ dependence.  With the SW approximation, we find 
\begin{eqnarray}
 b_{\ell_1 \ell_2 \ell_3, \rm SW}^{TT\mu, \rm const}
   \propto \ell_1^{-5/4} \ell_2^{-5/4} ~,
  \end{eqnarray}
rescaling $b_{\ell_1 \ell_2 \ell_3}^{TT\mu, d_0}$, $b_{\ell_1 \ell_2 \ell_3}^{TT\mu, d_2}$ and $b_{\ell_1 \ell_2 \ell_3}^{TT\mu, g_{\rm NL}}$ in Fig.~\ref{fig:blllTTmu_3D}.


\bibliography{paper}
\end{document}